\documentclass[12pt]{article}

\usepackage{graphicx}
\usepackage{amsmath}
\usepackage{amssymb}
\usepackage{here}
\usepackage{bm}

\allowdisplaybreaks

\begin{document}

\baselineskip=18.8pt plus 0.2pt minus 0.1pt

\makeatletter

\@addtoreset{equation}{section}
\renewcommand{\theequation}{\thesection.\arabic{equation}}
\renewcommand{\thefootnote}{\fnsymbol{footnote}}
\newcommand{\beq}{\begin{equation}}
\newcommand{\eeq}{\end{equation}}
\newcommand{\bea}{\begin{eqnarray}}
\newcommand{\eea}{\end{eqnarray}}
\newcommand{\nn}{\nonumber\\}
\newcommand{\hs}[1]{\hspace{#1}}
\newcommand{\vs}[1]{\vspace{#1}}
\newcommand{\Half}{\frac{1}{2}}
\newcommand{\p}{\partial}
\newcommand{\ol}{\overline}
\newcommand{\wt}[1]{\widetilde{#1}}
\newcommand{\ap}{\alpha'}
\newcommand{\bra}[1]{\left\langle  #1 \right\vert }
\newcommand{\ket}[1]{\left\vert #1 \right\rangle }
\newcommand{\vev}[1]{\left\langle  #1 \right\rangle }
\newcommand{\ul}[1]{\underline{#1}}

\makeatother

\begin{titlepage}
\title{
\vspace{1cm}
On the Relationship between Discrete and Continuous Energy Spectra of SU($N$) Supermembrane Matrix Model
}
\author{Yoji Michishita
\thanks{
{\tt michishita@edu.kagoshima-u.ac.jp}
}
\\[7pt]
{\it Department of Physics, Faculty of Education, Kagoshima University}\\
{\it Kagoshima, 890-0065, Japan}
}

\date{\normalsize August, 2013}
\maketitle
\thispagestyle{empty}

\begin{abstract}
\normalsize
It has been known that SU($N$) supermembrane matrix model has continuous energy spectrum, 
and it has also been conjectured that it has a normalizable energy eigenstate. 
Assuming that there exists a normalizable energy eigenstate for each $N$, 
we show that there exists a branch of continuous energy spectrum for each partition of $N$.
\end{abstract}
\end{titlepage}

\clearpage
\section{Introduction}

The SU($N$) supermembrane matrix quantum mechanics, which is
obtained from the dimensional reduction of $(9+1)\,$D SYM to $(0+1)\,$D,
describes low energy physics of $N$ D0-branes in type IIA string theory,
gives a regularization of M2-brane effective action in M-theory\cite{whn88},
and is expected to describe discrete light cone quantized M-theory\cite{bfss96, s97}.

To know more about string theory and M-theory, it is important to study this
system quantum mechanically, especially to study the structure of energy spectrum. In \cite{wln89},
it has been shown that this system has continuous spectrum.
Therefore most of eigenstates are expected
to be nonnormalizable. However this fact does not forbid existence of normalizable energy eigenstates,
and indeed the following conjecture has been made: There exists a unique normalizable zero 
energy eigenstate. This conjecture is natural in the viewpoint of D0-brane physics: 
if we uplift 10D type IIA string theory to 11D M-theory, a single D0-brane is regarded as
a Kaluza-Klein(KK) mode of one momentum unit along 11-th direction. $N$ D0-branes correspond to
$N$ KK modes of one momentum unit, and a single KK mode of $N$ times the momentum unit
can be given as a threshold bound state of $N$ KK modes of one momentum unit.

Then we are naturally led to the following description of the continuous spectrum:
Let us partition $N$ into positive integers $N_\mu$: $N=\sum_{\mu=1}^{n_b} N_\mu$.
$N$ D0-branes can form $n_b$ bound states which consist of $N_\mu$ D0-branes respectively.
This can be regarded as a $n_b$ particle state, and if those particles are far apart from
each other they behave as free particles. Therefore it gives a branch of continuous spectrum.

The purpose of this paper is to make the above description of the continuous spectrum 
more rigorous.
We do not inquire into the spectrum of normalizable states, but just assume that there exists
a normalizable energy eigenstate in the SU($N$) quantum mechanics for each $N$,
and the wavefunctions of those normalizable states decay sufficiently fast at infinity.
Extending the argument given in \cite{wln89}, we shall show that for each partition of $N$
there is a branch of continuous spectrum. The argument in \cite{wln89} corresponds to the case
where $N_\mu=1$ for any $\mu$. 

This paper is organized as follows. After summarizing notation in Section 2,
we shall show in Section 3 that using normalizable energy eigenstates $\psi_{(\mu)}$
of energy $E_{(\mu)}$ taken from SU($N_\mu$) subsystems, we can construct a smooth gauge invariant
function $\psi_{t,L}$ with two parameters $t$ and $L$ which has the following property:
\begin{quote}
For any $E\in[0,\infty)$ and $\epsilon>0$, there exist $L_0$ and $t_0(L)$ such that
\begin{equation*}
\forall L>L_0~~\text{and}~~ \forall t>t_0(L),\quad
||\psi_{t,L}||=1~\text{and}~ \Big|\Big|\Big(H-E-\sum_\mu E_{(\mu)}\Big)\psi_{t,L}\Big|\Big|<\epsilon.
\end{equation*}
\end{quote}
where $H$ is the Hamiltonian of the SU($N$) quantum mechanics.
Roughly speaking, $L$ restricts the size of the normalizable bound states $\psi_{(\mu)}$,
and $t$ is the distances between them. This fact means that there are branches of 
continuous energy spectrum of ranges $\Big[\sum_\mu E_{(\mu)}, \infty\Big)$.

In Section 4, in order to ensure that the above branches are independent of each other,
we shall show that inner products of $\psi_{t,L}$ corresponding to different partitions of $N$, 
or corresponding to different eigenstates of SU($N_\mu$) subsystems, can be taken arbitrarily small i.e.
\begin{quote}
For any $\epsilon>0$, there exist $L_0$ and $t_0(L, L')$ such that
\begin{equation*}
\forall L,L'>L_0 ~~\text{and}~~ \forall t,t'>t_0(L,L'),\quad
\big|\vev{\psi_{t,L}, \psi'_{t',L'}}\big|<\epsilon,
\end{equation*}
if $\psi_{t,L}$ and $\psi'_{t',L'}$ correspond to different partitions of $N$
or different eigenstates of SU($N_\mu$) subsystems.
\end{quote}

Section 5 contains some discussions.
In Appendix A we collect information on group theory necessary for the analysis.
In Appendix B we define some auxiliary functions used for defining $\psi_{t,L}$, and discuss
some of their properties. In Appendix C we discuss smoothness of eigenvalues and matrices used for
defining $\psi_{t,L}$.

\section{Preliminaries}

In this section we first have a quick review of the setup used in \cite{wln89},
and then we extend it to the one suitable for our purpose.
\subsection{Diagonally gauge fixed description of SU($N$) supermembrane matrix model}

The SU($N$) supermembrane matrix quantum mechanics is described
by Grassmann even hermitian traceless matrices $X^I$ and $X^9$, and
Grassmann odd hermitian traceless matrices $\theta_\alpha$, where
$(I,9)=(1,\dots,8,9)$ is an SO(9) vector index, and $\alpha=1,2,\dots,16$ is an SO(9) spinor index.
Gamma matrices $(\gamma^I)^{\alpha\beta}$ and $(\gamma^9)^{\alpha\beta}$ are real and symmetric, satisfying
\beq
\{\gamma^I,\gamma^J\}=2\delta^{IJ},\quad \{\gamma^I,\gamma^9\}=0,\quad (\gamma^9)^2=1.
\eeq
Using the basis which diagonalizes $\gamma^9$, $\alpha$ splits into $\alpha'$ and $\alpha''$ as follows:
\beq
\gamma^9=\begin{pmatrix} \delta_{\alpha'\beta'} & \\ & -\delta_{\alpha''\beta''}
\end{pmatrix},\quad
\gamma^I=\begin{pmatrix} & (\gamma^I)_{\alpha'\beta''} \\ (\gamma^I)_{\alpha''\beta'} &
\end{pmatrix}.
\eeq
We describe SU($N$) Lie algebra with a Cartan-Weyl basis 
$\{h_m,E_{ij};\;m=1,2,\dots,N-1,\; i,j=1,2,\dots,N,\;i\not\neq j\}$
(For notation about SU($N$) see Appendix A). $\theta_\alpha$ are expanded as
\beq
\theta_\alpha=\theta_\alpha^mh_m+\theta_\alpha^{(ij)}E_{ij}.
\eeq
Here and in the following, $(i,j)$ component of $\theta_\alpha$ is denoted by $\theta_\alpha^{(ij)}$.
Unless otherwise stated, when a pair of indices $(ij)$ is repeated it implies summation
$\sum_{i,j,i\not\neq j}$. 
Independent degrees of freedom of the diagonal components $\theta_\alpha^{(ii)}$ are given by
$\theta_\alpha^m$.
The nonzero anticommutation relations of $\theta_\alpha$ are
\beq
\{\theta_\alpha^m,\theta_\beta^n\}=\delta_{\alpha\beta}\delta^{mn},\quad
\{\theta_\alpha^{(ij)},\theta_\beta^{(kl)}\}=\delta_{\alpha\beta}\delta^{il}\delta^{jk}.
\eeq
Note that $(\theta_\alpha^{(ij)})^\dagger=\theta_\alpha^{(ji)}$.
$X^9$ and $X^I$, and their conjugate momenta $\Pi^9$ and $\Pi^I$
can be expanded analogously, and their nonzero commutation relations are 
\beq
[X^{Im}, \Pi^J_n]=i\delta^{IJ}\delta^{mn},\quad
[X^{I(ij)}, \Pi^J_{(kl)}]=i\delta^{IJ}\delta^i_k\delta^j_l,
\eeq
and analogously for $X^9$ and $\Pi^9$. Then the momenta are regarded as
$\Pi_m^I=-i\frac{\p}{\p X^{Im}}$, $\Pi_{(ij)}^I=-i\frac{\p}{\p X^{I(ij)}}$, 
and analogously for $\Pi^9$.

SU($N$) gauge invariant Hamiltonian $H$ of this quantum mechanics is
\bea
H & = & \text{tr}\Big[\frac{1}{2}\Pi^I\Pi^I+\frac{1}{2}\Pi^9\Pi^9
-\frac{1}{4}[X^I,X^J]^2-\frac{1}{2}[X^9,X^I]^2
\nn & &
+\frac{1}{2}\theta\gamma^9[X^9,\theta]
+\frac{1}{2}\theta\gamma^I[X^I,\theta]\Big].
\eea
Generators of the gauge transformation $G$ is decomposed into $X^9$ independent part $\hat{G}$
and $X^9$ dependent part $G^9$: $G=\hat{G}+G^9$, where
\bea
G^9_m & = & (h_m^i-h_m^j)[iX^{9(ij)}\Pi^9_{(ij)}],
\\
G^9_{(ij)} & = & (h_m^i-h_m^j)[iX^{9(ji)}\Pi^9_m-iX^{9m}\Pi^9_{(ij)}]
\nn & &
+i\sum_{k\not\neq i,j}[X^{9(jk)}\Pi^9_{(ik)}-X^{9(ki)}\Pi^9_{(kj)}],
\\\hat{G}_m & = & (h_m^i-h_m^j)[iX^{I(ij)}\Pi^I_{(ij)}
+\frac{1}{2}\theta_\alpha^{(ij)}\theta_\alpha^{(ji)}],
\\
\hat{G}_{(ij)} & = & (h_m^i-h_m^j)[iX^{I(ji)}\Pi^I_m-iX^{Im}\Pi^I_{(ij)}
+\theta_\alpha^{(ji)}\theta_\alpha^m]
\nn & &
+i\sum_{k\not\neq i,j}[X^{I(jk)}\Pi^I_{(ik)}-X^{I(ki)}\Pi^I_{(kj)}
-i\theta_\alpha^{(jk)}\theta_\alpha^{(ki)}].
\eea
$X^I$ dependent part and $\theta_\alpha$ dependent part of $\hat{G}$ are denoted by 
$\hat{G}^B$ and $\hat{G}^F$ respectively.
Infinitesimal gauge transformation is given by $\delta X^I=i[\epsilon, X^I]$ etc.
By an appropriate SU($N$) gauge transformation, $X^9$ can be diagonalized. 
Diagonal parts are denoted by $Z$ and $Z^I$,
and nondiagonal parts are denoted by $Y^I$:
\bea
X^9 & = & Z^mh_m, \\
X^I & = & \sum_{i,j}X^{I(ij)}E_{ij}=Z^{Im}h_m+Y^{I(ij)}E_{ij}.
\eea
For convenience we define $Y^{I(ii)}$ as $Y^{I(ii)}=0$.
Diagonal elements in $Z=\text{diag}(Z_1,Z_2,\dots,Z_N)$ are sorted into the order
$Z_1\geq Z_2\geq\dots\geq Z_N$. Since we have no overall U(1) part, $\sum Z_i=0$.
Components $Z^m$ are related to $Z_i$ by $Z^m=h_m^iZ_i$.
Analogous relations hold for $Z^{Im}$ and $Z^I_i$.

In general, a gauge invariant wavefunction $\psi(X^9, X^I)$ is reduced to the gauge fixed 
function $\hat{\psi}(Z,X^I)$ defined as
\beq
\hat{\psi}(Z, X^I)=C^{1/2}\cdot\prod_{i<j}(Z_i-Z_j)\cdot\psi(Z, X^I).
\eeq
where the factor $\prod_{i<j}(Z_i-Z_j)$ is Vandermonde determinant for $X^9$.
$C$ is a certain constant basically equal to the volume of SU($N$)$/K_0$, where
$K_0$ is the Cartan subgroup of SU($N$).
These factors are introduced in order for \eqref{ipip} to hold. 
This $\hat{\psi}$ is invariant under the action of $K_0$, and
is defined in the region $p=\{Z_i|Z_i\geq Z_j (i<j)\}$.
Conversely, if we have a function $\hat{\psi}$ invariant under the action of $K_0$ and defined in $p$,
we can easily reconstruct the original gauge invariant wavefunction $\psi$:
\beq
\psi(X^9,X^I)=C^{-1/2}\cdot\prod_{i<j}(Z_i-Z_j)^{-1}\cdot V_F(U)\hat{\psi}(Z, U^{-1}X^IU),
\eeq
where $Z=U^{-1}X^9U$ and $V_F(U)$ is the gauge transformation operator for fermion part corresponding to $U$.

The action of the kinetic operator in the Hamiltonian is translated to the action on $\hat{\psi}$ as
\beq
\text{tr}\Big[\Pi^9\Pi^9\Big]\psi\Big|_{X^9=Z}
=C^{-1/2}\cdot\prod_{k<l}(Z_k-Z_l)^{-1}\cdot\Big[-\left(\frac{\p}{\p Z^m}\right)^2
+(Z_i-Z_j)^{-2}\hat{G}_{(ij)}\hat{G}_{(ji)}\Big]\hat{\psi},
\label{redkin}
\eeq
and the inner products for gauge invariant functions
\beq
\vev{\psi_1, \psi_2}\equiv\int dX^9dX^I \psi_1^\dagger(X^9,X^I)\psi_2(X^9,X^I),
\eeq
are reduced to those for corresponding gauge fixed functions
\beq
\vev{\hat{\psi}_1, \hat{\psi}_2}\equiv\int_p dZdZ^IdY^I\hat{\psi}_1^\dagger(Z,Z^I,Y^I)\hat{\psi}_2(Z,Z^I,Y^I),
\eeq
defined so that
\beq
\vev{\psi_1, \psi_2}=\vev{\hat{\psi}_1, \hat{\psi}_2}.
\label{ipip}
\eeq
The norm $||\psi||=||\hat{\psi}||$ is given by
$||\psi||^2=||\hat{\psi}||^2=\vev{\psi, \psi}=\vev{\hat{\psi}, \hat{\psi}}$.

\subsection{Block decomposed description of SU($N$) supermembrane matrix model}

The description of the quantum mechanics in the previous subsection
is used in \cite{wln89} to construct a trial wavefunction for showing that 
this system has a branch of continuous spectrum. Let us extend this description 
to show the existence of other branches.
First, we take a set of positive integers $\{N_\mu\}$ satisfying $N=\sum_{\mu=1}^{n_b}N_\mu$, and
using it we decompose $N\times N$ matrices into $n_b\times n_b$ blocks. These blocks are indexed
by $\mu$, and the size of the $(\mu,\nu)$ block is $N_\mu\times N_\nu$.
Elements of matrices in $\mu$-th block is indexed by $i_\mu$. Note that $E_{i_\mu j_\mu}$ in SU($N$) Lie algebra
can be regarded as a generator of SU($N_\mu$) Lie algebra. However $h_{m_\mu}$ cannot be regarded as
a generator of SU($N_\mu$) Lie algebra. Elements in SU($N_\mu$) Cartan subalgebra are denoted by
$h_{(\mu)m_\mu}$.

Assume that nonzero components of $X^9$ are only in the diagonal blocks:
\beq
X^9=X^9_D\equiv\begin{pmatrix} X_{(1)} & & & \\
 & X_{(2)} & & \\
 & & \ddots & \\
 & & & X_{(n_b)}
\end{pmatrix},
\label{bdx9}
\eeq
where $X_{(\mu)}$ are $N_\mu\times N_\mu$ hermitian matrices, which are not necessarily traceless.
$X^I_{(\mu)}$ are also defined analogously: 
$X^{I(i_\mu j_\mu)}=X^{I(i_\mu j_\mu)}_{(\mu)}, X^{I(i_\mu j_\nu)}=Y^{I(i_\mu j_\nu)}~(\mu\not\neq\nu)$.
$X_{(\mu)}$ are further decomposed into diagonal and nondiagonal part:
$X_{(\mu)}=Z_{(\mu)}+Y_{(\mu)}$. Elements in $Z_{(\mu)}$ consist of
U(1) part $\Lambda_{(\mu)}$ and SU($N_\mu$) part $\lambda_{i_\mu}$:
\beq
Z_{(\mu)}^{(i_\mu i_\mu)} = \Lambda_{(\mu)}+\lambda_{i_\mu}
,\quad
\sum_{i_\mu}\lambda_{i_\mu}=0.
\eeq
For a block of $N_\mu=1$, $Z_{(\mu)}^{(i_\mu i_\mu)}=\Lambda_{(\mu)}$.
$\lambda^{m_\mu}$ is defined by $\lambda_{i_\mu}=h_{(\mu)m_\mu}^{i_\mu}\lambda^{m_\mu}$.
Since we have no overall U(1) part, $\sum_\mu N_\mu\Lambda_{(\mu)}=0$.
$\Lambda^I_{(\mu)}$, $\lambda^I_{i_\mu}$
and $\lambda^{Im_\mu}$ are defined analogously.

The residual gauge transformation which does not change the block diagonal form of
$X^9=X^9_D$ is given by $X^9\rightarrow uX^9u^{-1}$ etc. where
\beq
u=\begin{pmatrix} u_{(1)} & & & \\
 & u_{(2)} & & \\
 & & \ddots & \\
 & & & u_{(n_b)}
\end{pmatrix},
\label{residualgauge}
\eeq
and $u_{(\mu)}$ are $N_\mu\times N_\mu$ unitary matrices satisfying $\prod_\mu\det(u_{(\mu)})=1$.
These form a subgroup $K$ of SU($N$). By $K$, $X_{(\mu)}$ can further be diagonalized.
The eigenvalues of $X_{(\mu)}$ and its traceless part $X_{(\mu)}-\Lambda_{(\mu)}I_{(\mu)}$
are denoted by $Z_{i_\mu}$ and $z_{i_\mu}$ respectively, where $I_{(\mu)}$ 
is the $N_\mu\times N_\mu$ unit matrix. Then $Z_{i_\mu}=\Lambda_{(\mu)}+z_{i_\mu}$.

The following $n_b\times n_b$ matrices can be regarded as elements of SU($n_b$) Lie algebra: 
\beq
\Lambda=\begin{pmatrix} N_1\Lambda_{(1)} & & & \\
 & N_2\Lambda_{(2)} & & \\
 & & \ddots & \\
 & & & N_{n_b}\Lambda_{(n_b)}
\end{pmatrix},\quad
\Lambda^I=\begin{pmatrix} N_1\Lambda^I_{(1)} & & & \\
 & N_2\Lambda^I_{(2)} & & \\
 & & \ddots & \\
 & & & N_{n_b}\Lambda^I_{(n_b)}
\end{pmatrix},
\eeq
and these can be expanded by elements $h_{(0)M}$ of Cartan subalgebra of SU($n_b$):
\beq
\Lambda_{(\mu)}=\frac{1}{N_\mu}h_{(0)M}^\mu\Lambda^M,\quad
\Lambda^I_{(\mu)}=\frac{1}{N_\mu}h_{(0)M}^\mu\Lambda^{IM}.
\eeq
If we use $\lambda^{m_\mu}$ and $\Lambda^M$ as independent variables instead of $X^{9m}$,
the derivative operator $\frac{\p}{\p X^{9m}}$ is expressed as
\beq
\frac{\p}{\p X^{9m}}=\sum_\mu\sum_{i_\mu}h_m^{i_\mu}
\Big[h_{(\mu)m_\mu}^{i_\mu}\frac{\p}{\p\lambda^{m_\mu}}
+h_{(0)M}^\mu\frac{\p}{\p\Lambda^M}\Big],
\eeq
and analogously for $X^{Im}$. Therefore
\beq
\left(\frac{\p}{\p X^{9m}}\right)^2 = 
\left(\frac{\p}{\p\lambda^{m_\mu}}\right)^2+Q_{MN}\frac{\p}{\p\Lambda^M}\frac{\p}{\p\Lambda^N},
\eeq
where $Q_{MN}=h_{(0)M}^\mu{\cal N}_{\mu\nu}h_{(0)N}^\nu$ and
${\cal N}_{\mu\nu}=N_\mu\delta^{\mu\nu}-\frac{N_\mu N_\nu}{N}$.
The following can be used to evaluate terms in $\hat{G}_{(i_\mu j_\nu)}$:
\bea
(h_m^{i_\mu}-h_m^{j_\nu})\frac{\p}{\p X^{9m}}
 & =  & h_{(\mu)m_\mu}^{i_\mu}\frac{\p}{\p\lambda^{m_\mu}}
-h_{(\nu)m_\nu}^{j_\nu}\frac{\p}{\p\lambda^{m_\nu}}
+(h_{(0)M}^\mu-h_{(0)M}^\nu)\frac{\p}{\p\Lambda^M}.
\eea
The commutator of $X^9_D$ and $E_{i_\mu j_\nu}$ is given as
\beq
{}[X^9_D,E_{i_\mu j_\nu}]=(z_{\mu\nu})_{(i_\mu j_\nu)}{}^{(k_\mu l_\nu)}E_{k_\mu l_\nu},
\eeq
where
\bea
(z_{\mu\nu})_{(i_\mu j_\nu)}{}^{(k_\mu l_\nu)}
 & = & 
 X_{(\mu)}^{(k_\mu i_\mu)}\delta_{j_\nu}{}^{l_\nu}
-\delta^{k_\mu}{}_{i_\mu} X_{(\nu)}^{(j_\nu l_\nu)},
\eea
Analogously we define
\bea
(z^I_{\mu\nu})_{(i_\mu j_\nu)}{}^{(k_\mu l_\nu)}
 & = & 
 X^{I(k_\mu i_\mu)}_{(\mu)}\delta_{j_\nu}{}^{l_\nu}
-\delta^{k_\mu}{}_{i_\mu} X^{I(j_\nu l_\nu)}_{(\nu)}.
\eea
Since $X_{(\mu)}$ are hermitian, $z_{\mu\nu}$ are also hermitian, and are diagonalizable
by unitary matrices. When $X_{(\mu)}$ and $X_{(\nu)}$ have no common eigenvalue, 
$z_{\mu\nu}$ is invertible.

If we have a function $\hat{\psi}(X^9_D,X^I)$ which is invariant under $K$:
\beq
\hat{\psi}(X^9_D,X^I)=V_F(u^{-1})\hat{\psi}(uX^9_Du^{-1}, uX^Iu^{-1}),
\eeq
then we can define an SU($N$) invariant function of $X^9$ which is not necessarily
block diagonal:
\beq
\psi(X^9,X^I)=V^{-1/2}\Delta^{-1}V_F(U)\hat{\psi}(X^9_D, U^{-1}X^IU),
\eeq
where $\Delta=\prod_{\mu<\nu}\det(z_{\mu\nu})$, $V$ is equal to 
$\sqrt{\frac{N}{n_bN_1\dots N_{n_b}}}$ times the volume of SU($N$)$/K$,
and $U$ is an $N\times N$ unitary matrix which block diagonalizes
$X^9$: $X_D^9=U^{-1}X^9U$, in such a way that $Z_{i_\mu}\geq Z_{j_\nu}$ for $\mu<\nu$. 
Such a unitary matrix always exists, and elements of $U$ and $X_D$ are
smooth functions of the elements of $X^9$ when any pair of two different blocks
$X_{(\mu)}$ and $X_{(\nu)}$ have no common eigenvalue (see Appendix C for details). 
If some blocks have a common eigenvalue, then elements of $U$ and $X^9_D$
are not smooth. However in the following we consider functions
which are nonzero only when eigenvalues of different blocks are far apart from each other.
In this case $\psi(X^9,X^I)$ is smooth when $\hat{\psi}(X^9_D,X^I)$ is smooth.

Conversely, $\hat{\psi}(X^9_D,X^I)$ can be obtained from $\psi(X^9, X^I)$:
\beq
\hat{\psi}(X^9_D,X^I)=V^{1/2}\Delta\psi(X^9_D, X^I),
\eeq
and the integration measure $dX^9\cdot dX^I$ for $\psi(X^9, X^I)$ is reduced to
that for $\hat{\psi}(X^9_D,X^I)$ as follows:
\bea
dX^9\cdot dX^I & \equiv & \prod_mdX^{9m}\cdot\prod_{i\not\neq j}dX^{9(ij)}
 \cdot\prod_{I,m}dX^{Im}\cdot\prod_{I,i\not\neq j}dX^{I(ij)}
\nn \rightarrow  V\Delta^2 dX_D^9\cdot dX^I 
& \equiv &
 V\Delta^2 \prod_Md\Lambda^M\cdot\prod_{m_\mu}d\lambda^{m_\mu}
\cdot\prod_{i_\mu\not\neq j_\mu}dY_{(\mu)}^{(i_\mu j_\mu)} \cdot dX^I.
\eea
Then inner products for $\psi(X^9, X^I)$ are reduced to those for $\hat{\psi}(X^9_D,X^I)$:
\beq
\vev{\psi_1, \psi_2}=\vev{\hat{\psi}_1, \hat{\psi}_2},
\eeq
where
\bea
\vev{\psi_1, \psi_2} & \equiv & \int dX^9dX^I \psi_1^\dagger(X^9,X^I)\psi_2(X^9,X^I),
\\
\vev{\hat{\psi}_1, \hat{\psi}_2} & \equiv & \int_P dX^9_DdX^I
\hat{\psi}_1^\dagger(X^9_D,X^I)\hat{\psi}_2(X^9_D,X^I),
\eea
and $P=\{X_D^9~|~Z_{i_\mu}\geq Z_{j_\nu}~(\mu<\nu)\}$.
In this region $X_{(\mu)}$
cannot range over the entire space of $N_\mu\times N_\mu$ hermitian matrices.
However in the following we take only such integrands that their supports are 
compact subsets of the interior of $P$. So we can extend the range of 
$X_{(\mu)}$ to that of the entire $N_\mu\times N_\mu$ hermitian matrices.
The norm $||\psi||=||\hat{\psi}||$ is defined by
$||\psi||^2=||\hat{\psi}||^2=\vev{\psi, \psi}=\vev{\hat{\psi}, \hat{\psi}}$.

The action of $\Pi^9_{(i_\mu j_\nu)}$ on $\psi(X^9, X^I)$ is reduced to
\beq
\Pi^9_{(i_\mu j_\nu)}\psi(X^9, X^I)\Big|_{X^9=X^9_D}
=-i(z_{\mu\nu}^{-1})_{(i_\mu j_\nu)}{}^{(k_\mu l_\nu)}\hat{G}_{(k_\mu l_\nu)}\psi(X^9_D, X^I).
\eeq
Using this and the following property,
\beq
\Bigg[\left(\frac{\p}{\p\lambda^{m_\mu}}\right)^2
+Q_{MN}\frac{\p}{\p\Lambda^M}\frac{\p}{\p\Lambda^N}
+\frac{\p}{\p Y_{(\mu)}^{(i_\mu j_\mu)}}\frac{\p}{\p Y_{(\mu)}^{(j_\mu i_\mu)}}\Bigg]
\Delta=0,
\eeq
we can show that the kinetic operator $\text{tr}[\Pi^9\Pi^9]$ in the Hamiltonian acts 
on $\psi(X^9_D,X^I)$ as
\bea
\text{tr}[\Pi^9\Pi^9]
 & = & \Delta^{-1}\Big[
-\left(\frac{\p}{\p\lambda^{m_\mu}}\right)^2
-Q_{MN}\frac{\p}{\p\Lambda^M}\frac{\p}{\p\Lambda^N}
-\frac{\p}{\p Y_{(\mu)}^{(i_\mu j_\mu)}}\frac{\p}{\p Y_{(\mu)}^{(j_\mu i_\mu)}}
\nn & & 
+2\sum_{\mu<\nu}(z_{\mu\nu}^{-2})_{(i_\mu j_\nu)}{}^{(k_\mu l_\nu)}
\hat{G}_{(k_\mu l_\nu)}\hat{G}_{(j_\nu i_\mu)}
\Big]\Delta.
\eea

Then the reduced Hamiltonian $\hat{H}$ defined by
\beq
H\psi(X^9,X^I)\Big|_{X^9=X^9_D}=V^{-1/2}\Delta^{-1}\hat{H}\hat{\psi}(X^9_D,X^I),
\eeq
is decomposed as follows:
\beq
\hat{H}=\sum_\mu H_{(\mu)}+H_1+\sum_{\mu<\nu}H_2^{\mu\nu}+\sum_{\mu<\nu}H_3^{\mu\nu}+H_4,
\eeq
and the definitions of terms in the above are given as follows.
$H_{(\mu)}$ is the SU($N_\mu$) Hamiltonian for $\mu$-th diagonal block:
\bea
H_{(\mu)} & = & -\frac{1}{2}\Big[\left(\frac{\p}{\p\lambda^{m_\mu}}\right)^2
+\left(\frac{\p}{\p\lambda^{Im_\mu}}\right)^2
+\frac{\p}{\p Y_{(\mu)}^{(i_\mu j_\mu)}}\frac{\p}{\p Y_{(\mu)}^{(j_\mu i_\mu)}}
+\frac{\p}{\p Y_{(\mu)}^{I(i_\mu j_\mu)}}\frac{\p}{\p Y_{(\mu)}^{I(j_\mu i_\mu)}}
\Big]
\nn & &
+\sum_{i_\mu,j_\mu,k_\mu}\theta_\alpha^{(i_\mu j_\mu)}
\big[X_{(\mu)}^{(j_\mu k_\mu)}(\gamma^9)^{\alpha\beta}
+X_{(\mu)}^{I(j_\mu k_\mu)}(\gamma^I)^{\alpha\beta}\big]
\theta_\beta^{(k_\mu i_\mu)}
\nn & & 
+\sum_{i_\mu,j_\mu,k_\mu,l_\mu}\Big[
X_{(\mu)}^{(i_\mu j_\mu)}X_{(\mu)}^{(j_\mu k_\mu)}X_{(\mu)}^{I(k_\mu l_\mu)}X_{(\mu)}^{I(l_\mu i_\mu)}
-X_{(\mu)}^{(i_\mu j_\mu)}X_{(\mu)}^{I(j_\mu k_\mu)}X_{(\mu)}^{(k_\mu l_\mu)}X_{(\mu)}^{I(l_\mu i_\mu)}
\nn & &
+\frac{1}{2}
\big(X_{(\mu)}^{I(i_\mu j_\mu)}X_{(\mu)}^{I(j_\mu k_\mu)}X_{(\mu)}^{J(k_\mu l_\mu)}X_{(\mu)}^{J(l_\mu i_\mu)}
-X_{(\mu)}^{I(i_\mu j_\mu)}X_{(\mu)}^{J(j_\mu k_\mu)}X_{(\mu)}^{I(k_\mu l_\mu)}X_{(\mu)}^{J(l_\mu i_\mu)}\big)
\Big],
\eea
and for $N_\mu=1$, we define $H_{(\mu)}$ as $H_{(\mu)}=0$.

$H_1$ is the free Hamiltonian for U(1) parts:
\beq
H_1 = -\frac{1}{2}Q_{MN}\Big[\frac{\p}{\p\Lambda^M}\frac{\p}{\p\Lambda^N}
+\frac{\p}{\p\Lambda^{IM}}\frac{\p}{\p\Lambda^{IN}}\Big].
\eeq

$H_2$ and $H_3$ are bosonic and fermionic "harmonic oscillator" parts:
\bea
H_2^{\mu\nu} & = & -\frac{\p}{\p Y^{I(i_\mu j_\nu)}}\frac{\p}{\p Y^{I(j_\nu i_\mu)}}
+Y^{I(i_\mu j_\nu)}(z_{\mu\nu}^2)_{(i_\mu j_\nu)}{}^{(k_\mu l_\nu)}Y^{I(l_\nu k_\mu)},
\\
H_3^{\mu\nu} & = & \theta_\alpha^{(l_\nu k_\mu)}
\big[(z_{\mu\nu})_{(i_\mu j_\nu)}{}^{(k_\mu l_\nu)}(\gamma^9)^{\alpha\beta}
+(z^I_{\mu\nu})_{(i_\mu j_\nu)}{}^{(k_\mu l_\nu)}(\gamma^I)^{\alpha\beta}
\big]\theta_\beta^{(i_\mu j_\nu)}.
\eea

$H_4$ is the rest of $\hat{H}$:
\bea
H_4 & = & 
\frac{1}{2}\sum_{\stackrel{\mbox{$\scriptstyle \mu,\nu,\lambda,\rho$}}
 {\mbox{$\scriptstyle \mu\not\neq\nu,\nu\not\neq\lambda,\lambda\not\neq\rho,\rho\not\neq\mu$}}}
\big[Y^{I(i_\mu j_\nu)}Y^{I(j_\nu k_\lambda)}Y^{J(k_\lambda l_\rho)}Y^{J(l_\rho i_\mu)}
-Y^{I(i_\mu j_\nu)}Y^{J(j_\nu k_\lambda)}Y^{I(k_\lambda l_\rho)}Y^{J(l_\rho i_\mu)}\big]
\nn & &
-\mathop{{\sum}'}_{\mu,\nu,\lambda}
(z_{\mu\nu}^I)_{(i_\mu j_\nu)}{}^{(k_\mu l_\nu)}Y^{J(i_\mu j_\nu)}
\big[Y^{I(l_\nu p_\lambda)}Y^{J(p_\lambda k_\mu)}-Y^{J(l_\nu p_\lambda)}Y^{I(p_\lambda k_\mu)}\big]
\nn & &
+\frac{1}{2}\mathop{{\sum}'}_{\mu,\nu}
 ((z_{\mu\nu}^I)^2)_{(i_\mu j_\nu)}{}^{(k_\mu l_\nu)}Y^{J(i_\mu j_\nu)}Y^{J(l_\nu k_\mu)}
+\frac{1}{2}\mathop{{\sum}'}_{\mu,\nu}
 (z_{\mu\nu}^Iz_{\mu\nu}^J)_{(i_\mu j_\nu)}{}^{(k_\mu l_\nu)}Y^{I(i_\mu j_\nu)}Y^{J(l_\nu k_\mu)}
\nn & &
-\mathop{{\sum}'}_{\mu,\nu}
 (z_{\mu\nu}^Iz_{\mu\nu}^J)_{(i_\mu j_\nu)}{}^{(k_\mu l_\nu)}Y^{J(i_\mu j_\nu)}Y^{I(l_\nu k_\mu)}
\nn & &
+\mathop{{\sum}'}_{\mu,\nu,\lambda}\theta_\alpha^{(j_\nu i_\mu)}(\gamma^I)^{\alpha\beta}
Y^{I(i_\mu k_\lambda)}\theta_\beta^{(k_\lambda j_\nu)}
\nn & &
+\mathop{{\sum}'}_{\mu,\lambda}\theta_\alpha^{(j_\mu i_\mu)}(\gamma^I)^{\alpha\beta}
Y^{I(i_\mu k_\lambda)}\theta_\beta^{(k_\lambda j_\mu)}
+\mathop{{\sum}'}_{\mu,\nu}\theta_\alpha^{(j_\nu i_\mu)}(\gamma^I)^{\alpha\beta}
Y^{I(i_\mu k_\nu)}\theta_\beta^{(k_\nu j_\nu)}
\nn & & 
-\frac{1}{2}\mathop{{\sum}'}_{\mu,\nu}(z_{\mu\nu}^{-2})_{(i_\mu j_\nu)}{}^{(k_\mu l_\nu)}
\hat{G}_{(k_\mu l_\nu)}\hat{G}_{(j_\nu i_\mu)},
\eea
where $\Sigma'$ implies summation which counts only the case where all the dummy indices are different.
$\hat{H}$ has the following property for any positive integer $q$:
\beq
\vev{\psi_1, H^q\psi_2}=\vev{\hat{\psi}_1, \hat{H}^q\hat{\psi}_2}
\eeq

\section{Construction of Trial Wavefunction}

It has been conjectured that there exists a unique normalizable zero energy eigenstate
in the SU($N$) supermembrane quantum mechanics.
In addition to it there may be excited normalizable states
(see e.g. \cite{s08}).
These normalizable states can be taken orthogonal to each other.
Here we only postulate that there exists at least one normalizable energy eigenstate for each $N$.
Let $\psi(X^9,X^I)$ be such a state with energy eigenvalue $E$. Then $H\psi=E\psi$, 
and since it is normalizable i.e. we can set $||\psi||=1$,
it decays sufficiently fast at infinity. Therefore we assume that
$\vev{\psi,P(X^9.X^I,\frac{\p}{\p X^9},\frac{\p}{\p X^I})\psi}$ are finite 
for any polynomial $P$ of $X^9$, $X^I$ and derivative operators of $X^9$ and $X^I$. 

\subsection{The trial wave function}

We take a set of normalizable energy eigenstates $\{\psi_{(\mu)}\}$ from 
each SU($N_\mu$) quantum mechanics which are regarded as subsystems of the entire SU($N$) system.
Energy eigenvalues of these states are denoted by $E_{(\mu)}$ i.e.
\beq
H_{(\mu)}\psi_{(\mu)}=E_{(\mu)}\psi_{(\mu)},\quad ||\psi_{(\mu)}||=1.
\eeq
For $N_\mu=1$, we define $E_{(\mu)}$ and $\psi_{(\mu)}$ as $E_{(\mu)}=0$ and $\psi_{(\mu)}=1$.
These are invariant under $K$.
Our goal in this section is to show the following fact using $\{\psi_{(\mu)}\}$:
It is possible to construct a function $\hat{\psi}_{t,L}$ with parameters $t$ and $L$ 
satisfying the following condition: 
$\hat{\psi}_{t,L}$ is smooth and invariant under $K$, and
for arbitrary nonnegative $E$ and positive $\epsilon$, there exist $L_0$ and $t_0$ such that
\beq
\forall L>L_0 ~~\text{and}~~ \forall t>t_0, \quad ||\hat{\psi}_{t,L}||=1 ~~\text{and}~~
\Big|\Big|\Big(\hat{H}-E-\sum_\mu E_{(\mu)}\Big)\hat{\psi}_{t,L}\Big|\Big|<\epsilon,
\label{theorem1}
\eeq
where $L_0$ depends on $E$ and $\epsilon$, and $t_0$ depends on $E$, $\epsilon$, and $L$.

Such a function is given as follows:
\beq
\hat{\psi}_{t,L}=\chi_{L,E}(\Lambda^M-tD^M, \Lambda^{IM},\theta^M_\alpha)\cdot
\Psi\cdot\Xi_B\cdot\Xi_F,
\eeq
\beq
\Psi=\prod_\mu\chi_{(\mu)L}\psi_{(\mu)},\quad
\Xi_B=\prod_{\mu<\nu}\xi_B^{\mu\nu},\quad
\Xi_F=\prod_{\mu<\nu}\xi_F^{\mu\nu},
\eeq
where $D$ is the following diagonal traceless $n_b\times n_b$ matrix:
\beq
D=\begin{pmatrix} N_1\left(\frac{\sum_\mu\mu N_\mu}{N}-1\right) & & & & \\
 & N_2\left(\frac{\sum_\mu\mu N_\mu}{N}-2\right) & & & \\
 & & & \ddots & \\
 & & & & N_{n_b}\left(\frac{\sum_\mu\mu N_\mu}{N}-n_b\right)
\end{pmatrix}.
\eeq
Definitions of the factors in $\hat{\psi}_{t,L}$ will be given in the following.
The factors have the following dependence on the variables:
\bea
\chi_{(\mu)L} & = & \chi_{(\mu)L}(\lambda^{m_\mu}, 
 \lambda^{Im_\mu},Y_{(\mu)}^{(i_\mu j_\mu)},Y_{(\mu)}^{I(i_\mu j_\mu)}),
\\
\psi_{(\mu)} & = & \psi_{(\mu)}(\lambda^{m_\mu}, \lambda^{Im_\mu},
 Y_{(\mu)}^{(i_\mu j_\mu)},Y_{(\mu)}^{I(i_\mu j_\mu)},
\theta_\alpha^{m_\mu}, \theta_\alpha^{(i_\mu j_\mu)}),
\\
\xi_B^{\mu\nu} & = & \xi_B^{\mu\nu}(z_{\mu\nu}, Y^{I(i_\mu j_\nu)}),
\\
\xi_F^{\mu\nu} & = & \xi_F^{\mu\nu}(z_{\mu\nu},z^I_{\mu\nu},\theta_\alpha^{(i_\mu j_\nu)}).
\eea
$\chi_{(\mu)L}$ have finite supports characterized by $L$, and are given in Appendix B.
$\chi_{L,E}(\Lambda^M,\Lambda^{IM},\theta^M_\alpha)$ is also given in Appendix B,
and consists of bosonic part dependent on $\Lambda^M$ and $\Lambda^{IM}$,
and fermionic part dependent on $\theta^M_\alpha$.
We do not specify this fermionic part and ignore it because it is not necessary in the following
(see \cite{pw97} for details of this part.).
$\chi_{L,E}(\Lambda^M,\Lambda^{IM},\theta^M_\alpha)$ also has a finite support 
$|\Lambda^MW_{MN}|\leq L$ and $|\Lambda^{IM}W_{MN}|\leq L$. 
Then $\chi_{L,E}(\Lambda^M-tD^M, \Lambda^{IM},\theta^M_\alpha)$
has the support $|\wt{\Lambda}^M|\leq L$ and $|\wt{\Lambda}^{IM}|\leq L$,
where $\wt{\Lambda}^M$ are defined by either of the following:
\bea
N_\mu\Lambda_{(\mu)} & = & N_\mu\left(\frac{\sum_\nu\nu N_\nu}{N}
-\mu\right)t+h_{(0)M}^\mu W_{MN}\wt{\Lambda}^N,
\\
\Lambda^M & = & t\sum_\mu h_{(0)M}^\mu N_\mu\left(\frac{\sum_\nu\nu N_\nu}{N}-\mu\right)
+W_{MN}\wt{\Lambda}^N,
\eea
and analogously for $\wt{\Lambda}^{IM}$. Then
\bea
Z_{i_\mu}-Z_{j_\nu}
 & = & \Lambda_{(\mu)}-\Lambda_{(\nu)}+z_{i_\mu}-z_{j_\nu}
\nn
 & = & (\nu-\mu)t+\left(\frac{h_{(0)M}^\mu}{N_\mu}-\frac{h_{(0)M}^\nu}{N_\nu}\right)W_{MN}\wt{\Lambda}^N
+z_{i_\mu}-z_{j_\nu},
\label{limmljn}
\eea
and, because in general eigenvalues are bounded by the norm of the matrices,
$|z_{i_\mu}|\leq |X_{(\mu)}-\Lambda_{(\mu)}I_{(\mu)}|\leq L+\ell$ in the support of $\chi_{(\mu)L}$.
Therefore, if we take $t$ much larger than $L$,
the factor $\chi_{L,E}(\Lambda^M-tD^M, \Lambda^{IM},\theta^M_\alpha)\prod_\mu\chi_{(\mu)L}$ enables us to
consider $\hat{\psi}_{t,L}$ only in the region where the eigenvalues of $X^9_D$ in different blocks 
are sufficiently far apart from each other i.e. $Z_{i_\mu}\gg Z_{j_\nu} (\mu<\nu)$, 
while those in the same blocks are relatively close to each other.
The parameter $t$ characterizes the distances between different blocks.

The factor $\Psi$ is basically given as the product of $\psi_{(\mu)}$, but 
in order to restrict their supports the additional factors $\chi_{(\mu)L}$ are included.
$\chi_{(\mu)L}$ are functions of $r_\mu$ (see Appendix B), and their supports are 
$r_\mu\leq L+\ell$. Since $r_\mu$ is invariant under $K$, $\chi_{(\mu)L}$ is also invariant.
Then $\chi_{(\mu)L}\psi_{(\mu)}$ is invariant, and is normalized: $||\chi_{(\mu)L}\psi_{(\mu)}||=1$.
Though $\Psi$ is intended for giving an eigenfunction of $H_{(\mu)}$, 
derivative operators in $H_4$ also act on $\Psi$.

\subsection{Definition of $\xi_B^{\mu\nu}$}

$\xi_B^{\mu\nu}$ is defined as the "ground state" of $H_2^{\mu\nu}$:
\beq
\xi_B^{\mu\nu}=
\Big[\det\Big(\frac{2}{\pi}z_{\mu\nu}\Big)\Big]^4\cdot
\exp\big[-Y^{I(i_\mu j_\nu)}(z_{\mu\nu})_{(i_\mu j_\nu)}{}^{(k_\mu l_\nu)}Y^{I(l_\nu k_\mu)}\big].
\eeq
This is invariant under the action of $K$, is normalized:
\beq
\int\prod_{I,i_\mu,j_\nu}dY^{I(i_\mu j_\nu)}dY^{I(j_\nu i_\mu)}(\xi_B^{\mu\nu})^2=1,
\eeq
and is an eigenfunction of $H_2$: $H_2^{\mu\nu}\xi_B^{\mu\nu}=
8\,\text{tr}(z_{\mu\nu})\xi_B^{\mu\nu}$. Note that  
the derivative operators in $H_{(\mu)}$, $H_1$ and $H_4$ also act on $\xi_B^{\mu\nu}$.

If a function $P(\Lambda^M, \lambda^{m_\mu}, Y^{I(i_\mu j_\nu)})$, which can contain
derivative operators of the arguments, satisfies the condition
\beq
\int
\prod dY^{I(i_\mu j_\nu)}
\Xi_B^\dagger P(\Lambda^M, \lambda^{m_\mu}, Y^{I(i_\mu j_\nu)})\Xi_B
=O(t^n) \quad (t\rightarrow\infty),
\eeq
we assign $P$ (or $P\Xi_B$) degree $n$. For example,
\begin{align}
Y^{I(i_\mu j_\nu)} & : -1/2, & \p/\p Y^{I(i_\mu j_\nu)} & : 1/2, \notag\\
z_{\mu\nu} & : 1, & \p/\p\lambda^{m_\mu} & : -1.
\end{align}
Terms of negative degree can be ignored when we compute inner products 
and take $t$ much larger than $L$,
as long as the integrations over other variables are convergent.

\subsection{Definition of $\xi_F^{\mu\nu}$}

Next we give the definition of $\xi_F^{\mu\nu}$ for $\mu<\nu$.
A hermitian matrix ${\cal M}$ is defined as
\beq
{\cal M}^{\alpha(k_\mu l_\nu)}{}_{\beta(i_\mu j_\nu)}
=(z_{\mu\nu})_{(i_\mu j_\nu)}{}^{(k_\mu l_\nu)}(\gamma^9)^{\alpha\beta}
+(z^I_{\mu\nu})_{(i_\mu j_\nu)}{}^{(k_\mu l_\nu)}(\gamma^I)^{\alpha\beta}.
\eeq
Then $H_3= \theta_\alpha^{(l_\nu k_\mu)}{\cal M}^{\alpha(k_\mu l_\nu)}{}_{\beta(i_\mu j_\nu)}
\theta_\beta^{(i_\mu j_\nu)}$.
Eigenvalues of ${\cal M}$ are real, and ${\cal M}$ can be diagonalized by a unitary matrix 
${\cal U}^A{}_{\alpha(i_\mu j_\nu)}$: 
${\cal U}{\cal M}{\cal U}^\dagger=\text{diag} (m_1,m_2,\dots m_{16N_\mu N_\nu})$.
${\cal U}$ is given by ${\cal U}^\dagger=(v_1, v_2, \dots, v_{16N_\mu N_\nu})$,
where $v_A$ are normalized eigenvectors determined by
\beq
{\cal M}^{\alpha(k_\mu l_\nu)}{}_{\beta(i_\mu j_\nu)}v^{\beta(i_\mu j_\nu)}_A
=m_Av^{\alpha(k_\mu l_\nu)}_A,\quad
(v_A)^\dagger\cdot v_A=1.
\eeq
It is easy to see that under the action of $K$ \eqref{residualgauge}, $m_A$ is invariant and
\bea
v^{\alpha(i_\mu j_\nu)}_A & \rightarrow & 
u_{(\mu)}{}^{i_\mu}{}_{k_\mu}(u^\dagger_{(\nu)})^{l_\nu}{}_{j_\nu}v^{\alpha(k_\mu l_\nu)}_A,
\\
{\cal U}^A{}_{\alpha(i_\mu j_\nu)} & \rightarrow & 
u_{(\nu)}{}^{j_\nu}{}_{l_\nu}(u^\dagger_{(\mu)})^{k_\mu}{}_{i_\mu}{\cal U}^A{}_{\alpha(k_\mu l_\nu)}.
\eea
${\cal M}/t$ depends on $z_{\mu\nu}/t$ and $z^I_{\mu\nu}/t$.
Therefore $\wt{m}_A\equiv m_A/t$, $v_A$ and ${\cal U}$ are functions of these variables:
\beq
m_A=t\wt{m}_A(z_{\mu\nu}/t,z^I_{\mu\nu}/t),\quad
v_A=v_A(z_{\mu\nu}/t,z^I_{\mu\nu}/t),\quad
{\cal U}={\cal U}(z_{\mu\nu}/t,z^I_{\mu\nu}/t).
\eeq
By diagonalizing $X_{(\mu)}$ and $X_{(\nu)}$:
\beq
U_{(\mu)}{}^{i_\mu}{}_{k_\mu}X_{(\mu)}^{(k_\mu l_\mu)}(U_{(\mu)}^\dagger)^{l_\mu}{}_{j_\mu}
=Z_{i_\mu}\delta^{i_\mu}{}_{j_\mu},\quad
U_{(\nu)}{}^{i_\nu}{}_{k_\nu}X_{(\nu)}^{(k_\nu l_\nu)}(U_{(\nu)}^\dagger)^{l_\nu}{}_{j_\nu}
=Z_{i_\nu}\delta^{i_\nu}{}_{j_\nu},
\eeq
${\cal M}|_{z^I_{\mu\nu}=0}$ can be diagonalized:
\bea
U_{(\mu)}{}^{k_\mu}{}_{r_\mu}(U_{(\nu)}^\dagger)^{s_\nu}{}_{l_\nu}
\left({\cal M}|_{z^I_{\mu\nu}=0}\right)^{\alpha(r_\mu s_\nu)}{}_{\beta(p_\mu q_\nu)}
(U_{(\mu)}^\dagger)^{p_\mu}{}_{i_\mu}U_{(\nu)}{}^{j_\nu}{}_{q_\nu}
\nn
=\delta^{k_\mu}_{i_\mu}\delta^{l_\nu}_{j_\nu}
\Big[
(Z_{i_\mu}-Z_{j_\nu})\left(\frac{1+\gamma^9}{2}\right)^{\alpha\beta}
-(Z_{i_\mu}-Z_{j_\nu})\left(\frac{1-\gamma^9}{2}\right)^{\alpha\beta}
\Big].
\eea
Therefore ${\cal M}|_{z^I_{\mu\nu}=0}$ has eightfold degenerate positive eigenvalues 
$Z_{i_\mu}-Z_{j_\nu}$ and negative eigenvalues $-(Z_{i_\mu}-Z_{j_\nu})$.
For nonzero $z^I_{\mu\nu}$, the eigenvalues receive corrections dependent on $z^I_{\mu\nu}$.
The eigenvalues approaching $Z_{i_\mu}-Z_{j_\nu}$ and
$-(Z_{i_\mu}-Z_{j_\nu})$ as $z^I_{\mu\nu}\rightarrow 0$ are denoted by indices 
$[\alpha'(i_\mu j_\nu)]$ and $[\alpha''(i_\mu j_\nu)]$ respectively (Note that eigenvalues
are continuous functions of elements of matrices). Then
\bea
m_{[\alpha'(i_\mu j_\nu)]}
& = & Z_{i_\mu}-Z_{j_\nu}+t\Delta m_{[\alpha'(i_\mu j_\nu)]}(z_{\mu\nu}/t, z^I_{\mu\nu}/t),
\\
m_{[\alpha''(i_\mu j_\nu)]}
& = & -(Z_{i_\mu}-Z_{j_\nu})+t\Delta m_{[\alpha''(i_\mu j_\nu)]}(z_{\mu\nu}/t, z^I_{\mu\nu}/t),
\eea
where the corrections $\Delta m_{[\alpha(i_\mu j_\nu)]}~(\alpha=(\alpha',\alpha''))$
satisfy the following:
\beq
\Delta m(z_{\mu\nu}/t, 0)_{[\alpha(i_\mu j_\nu)]}=0.
\eeq
In the support of $\chi_{L,E}(\Lambda^M-tD^M, \Lambda^{IM},\theta^M_\alpha)\prod_\mu\chi_{(\mu)L}$,
absolute values of the components of $z^I_{\mu\nu}/t$ are small if we take $t\gg L$,
and then from the continuity of the eigenvalues, $m_{[\alpha'(i_\mu j_\nu)]}$ are
positive and $m_{[\alpha''(i_\mu j_\nu)]}$ are negative.

Let us show that $\Delta m(z_{\mu\nu}, z^I_{\mu\nu})_{[\alpha(i_\mu j_\nu)]}$ are
actually functions of products of two elements of $z^I_{\mu\nu}$.
Using the basis which diagonalizes $\gamma^9$, the eigenvalue equation for ${\cal M}$ is expressed as
\bea
0 & = & \det({\cal M}-mI)
\nn
& = & \det\begin{pmatrix} \big[(z_{\mu\nu})_{(i_\mu j_\nu)}{}^{(k_\mu l_\nu)}
-m\delta^{k_\mu}_{i_\mu}\delta^{l_\nu}_{j_\nu}\big]\delta_{\alpha'\beta'}
& 
(z^I_{\mu\nu})_{(i_\mu j_\nu)}{}^{(k_\mu l_\nu)}(\gamma^I)_{\alpha'\beta''}
\\
(z^I_{\mu\nu})_{(i_\mu j_\nu)}{}^{(k_\mu l_\nu)}(\gamma^I)_{\alpha''\beta'}
&
-\big[(z_{\mu\nu})_{(i_\mu j_\nu)}{}^{(k_\mu l_\nu)}
+m\delta^{k_\mu}_{i_\mu}\delta^{l_\nu}_{j_\nu}\big]\delta_{\alpha''\beta''}
\end{pmatrix}.
\eea
Let us apply the formula 
$\det\bigl(\begin{smallmatrix} A & C \\ D & B\end{smallmatrix}\bigr)=\det(B)\det(A-CB^{-1}D)$
to evaluate the above determinant. Then we see that $C$ and $D$ contain $z^I_{\mu\nu}$ linearly,
and $\det(A-CB^{-1}D)$ depends on the products of two elements of $z^I_{\mu\nu}$.
Therefore the eigenvalue $m$, and $\Delta m_{[\alpha(i_\mu j_\nu)]}$ depend on them.
In the following $\Delta m_{[\alpha(i_\mu j_\nu)]}(z_{\mu\nu}/t, z^I_{\mu\nu}/t)$ is expressed as 
$\Delta m_{[\alpha(i_\mu j_\nu)]}(z_{\mu\nu}/t, (z^I_{\mu\nu})^2/t^2)$
schematically.

Each $\Delta m_{[\alpha''(i_\mu j_\nu)]}$ is not a smooth function of $z_{\mu\nu}/t$ and
$(z^I_{\mu\nu})^2/t^2$ when $m_{[\alpha''(i_\mu j_\nu)]}$ is degenerate, but
the sum of them is smooth (see Appendix C).
Then $\Delta E^{\mu\nu}_{0F}$ defined by
\beq
\sum_{\alpha'',i_\mu,j_\nu}\Delta m_{[\alpha''(i_\mu j_\nu)]}
 = \frac{(z^I_{\mu\nu})^2}{t^2}\Delta E^{\mu\nu}_{0F}(z_{\mu\nu}/t, (z^I_{\mu\nu})^2/t^2)
\label{dms22}
\eeq
is a smooth function. (The right hand side of \eqref{dms22} is a schematic
expression, and actually means a sum of terms in the form of 
$(\text{product of two components of}~z^I_{\mu\nu}/t)\times$(smooth function).
Equations containing $\Delta E^{\mu\nu}_{0F}$ in the following should be understood similarly.)

We define $\wt{\theta}_A$ as
$\wt{\theta}_A={\cal U}^A{}_{\alpha(i_\mu j_\nu)}\theta_\alpha^{(i_\mu j_\nu)}$.
These satisfy $\{(\wt{\theta}_A)^\dagger,\wt{\theta}_B\}=\delta_{AB}$ and are
invariant under the action of $K$. Then
\bea
H_3^{\mu\nu} & = & \sum_A m_A(\wt{\theta}_A)^\dagger\wt{\theta}_A
\nn & = &
\sum_{\alpha',i_\mu,j_\nu} m_{[\alpha'(i_\mu j_\nu)]}
(\wt{\theta}_{[\alpha'(i_\mu j_\nu)]})^\dagger\wt{\theta}_{[\alpha'(i_\mu j_\nu)]}
\nn & & 
+\sum_{\alpha'',i_\mu,j_\nu} (-m_{[\alpha''(i_\mu j_\nu)]})
\wt{\theta}_{[\alpha''(i_\mu j_\nu)]}(\wt{\theta}_{[\alpha''(i_\mu j_\nu)]})^\dagger
+\sum_{\alpha'',i_\mu,j_\nu} m_{[\alpha''(i_\mu j_\nu)]}.
\eea
The last term in the last line of the above is the "zero point energy" for $\wt{\theta}$:
\bea
\sum_{\alpha'',i_\mu,j_\nu} m_{[\alpha''(i_\mu j_\nu)]}
 & = & -8\sum_{i_\mu,j_\nu}(Z_{i_\mu}-Z_{j_\nu})
 +t\sum_{\alpha'',i_\mu,j_\nu}\Delta m_{[\alpha''(i_\mu j_\nu)]}
\nn
 & = & -8\,\text{tr}(z_{\mu\nu})
 +\frac{(z^I_{\mu\nu})^2}{t}\Delta E^{\mu\nu}_{0F}(z_{\mu\nu}/t, (z^I_{\mu\nu})^2/t^2).
\eea
$\xi_F^{\mu\nu}$ is defined as the "ground state" of $H_3^{\mu\nu}$:
\beq
\xi_F^{\mu\nu}=\prod_{\alpha'',i_\mu,j_\nu}(\wt{\theta}_{[\alpha''(i_\mu j_\nu)]})^\dagger\ket{0}_{\mu\nu},
\eeq
where the vacuum $\ket{0}_{\mu\nu}$ is defined by $\theta_{\alpha''}^{(i_\mu j_\nu)}\ket{0}_{\mu\nu}=0$.
$\xi_F^{\mu\nu}$ is normalized:
\beq
\vev{\xi_F^{\mu\nu}, \xi_F^{\mu\nu}}=1,
\eeq
and is an eigenfunction of $H_3$:
\beq
H_3^{\mu\nu}\xi_F^{\mu\nu}=\Big[-8\,\text{tr}(z_{\mu\nu})
 +\frac{(z^I_{\mu\nu})^2}{t}\Delta E^{\mu\nu}_{0F}(z_{\mu\nu}/t, (z^I_{\mu\nu})^2/t^2)\Big] \xi_F^{\mu\nu}.
\eeq
Note that the derivative operators in $H_{(\mu)}$, $H_1$ and $H_4$ also act on $\xi_F^{\mu\nu}$.
We also note that
\beq
(H_2^{\mu\nu}+H_3^{\mu\nu})(\xi_B^{\mu\nu}\xi_F^{\mu\nu})
=\frac{(z^I_{\mu\nu})^2}{t}\Delta E^{\mu\nu}_{0F}(z_{\mu\nu}/t, (z^I_{\mu\nu})^2/t^2)\xi_B^{\mu\nu}\xi_F^{\mu\nu}.
\eeq

One may wonder if $\xi_F^{\mu\nu}$ is a smooth function of $z_{\mu\nu}/t$ and $z^I_{\mu\nu}/t$,
because in general ${\cal U}$ is not smooth when $m_{[\alpha(i_\mu j_\nu)]}$ are degenerate.
Its smoothness can be proven as follows. ${\cal M}$ can be block diagonalized by a smooth 
unitary matrix $\wt{{\cal U}}$ in the region where absolute values of the components of 
$z^I_{\mu\nu}/t$ are small (see Appendix C):
\beq
\wt{{\cal U}}{\cal M}\wt{{\cal U}}^\dagger
=\begin{pmatrix} {\cal M}_+ & \\ & {\cal M}_- \end{pmatrix}.
\eeq
where ${\cal M}_+$ has eigenvalues 
$m_{[\alpha'(i_\mu j_\nu)]}$, and ${\cal M}_-$ has eigenvalues 
$m_{[\alpha''(i_\mu j_\nu)]}$.
This can be further diagonalized by a block
diagonal unitary matrix $\bigl(\begin{smallmatrix} {\cal U}_+ & \\ & {\cal U}_- \end{smallmatrix}\bigr)$
which is not necessarily smooth. Determinants of ${\cal U}_+$ and ${\cal U}_-$ can be taken to be 1.
These unitary matrices are related to ${\cal U}$ by
\beq
{\cal U}=\begin{pmatrix} {\cal U}_+ & \\ & {\cal U}_- \end{pmatrix}\wt{{\cal U}}.
\eeq
$\xi_F$ can be written in the following form:
\beq
\xi_F^{\mu\nu}=\frac{1}{n!}\epsilon_{A_1A_2\dots A_n}
 ({\cal U}^{A_1}{}_{\alpha_1(i_{1\mu} j_{1\nu})}\theta_{\alpha_1}^{(i_{1\mu} j_{1\nu})})^\dagger
 \dots
 ({\cal U}^{A_n}{}_{\alpha_n(i_{n\mu} j_{n\nu})}\theta_{\alpha_n}^{(i_{n\mu} j_{n\nu})})^\dagger
 \ket{0}_{\mu\nu},
\eeq
where $n=8N_\mu N_\nu$, and indices $A_i$ range over indices of type $[\alpha''(i_\mu j_\nu)]$.
Then
\bea
\xi_F^{\mu\nu} & = & \frac{1}{n!}\epsilon_{A_1A_2\dots A_n}
 ({\cal U}_-{}^{A_1}{}_{B_1})^\dagger
 \dots ({\cal U}_-{}^{A_n}{}_{B_n})^\dagger
\nn & & 
 (\wt{\cal U}^{B_1}{}_{\alpha_1(i_{1\mu} j_{1\nu})}\theta_{\alpha_1}^{(i_{1\mu} j_{1\nu})})^\dagger
 \dots
 (\wt{\cal U}^{B_n}{}_{\alpha_n(i_{n\mu} j_{n\nu})}\theta_{\alpha_n}^{(i_{n\mu} j_{n\nu})})^\dagger
 \ket{0}_{\mu\nu}
\nn & = &
\frac{1}{n!}\det({\cal U}_-^\dagger)\epsilon_{B_1B_2\dots B_n}
 (\wt{\cal U}^{B_1}{}_{\alpha_1(i_{1\mu} j_{1\nu})}\theta_{\alpha_1}^{(i_{1\mu} j_{1\nu})})^\dagger
 \dots
 (\wt{\cal U}^{B_n}{}_{\alpha_n(i_{n\mu} j_{n\nu})}\theta_{\alpha_n}^{(i_{n\mu} j_{n\nu})})^\dagger
 \ket{0}_{\mu\nu}
 \nn & = &
\frac{1}{n!}\epsilon_{B_1B_2\dots B_n}
 (\wt{\cal U}^{B_1}{}_{\alpha_1(i_{1\mu} j_{1\nu})}\theta_{\alpha_1}^{(i_{1\mu} j_{1\nu})})^\dagger
 \dots
 (\wt{\cal U}^{B_n}{}_{\alpha_n(i_{n\mu} j_{n\nu})}\theta_{\alpha_n}^{(i_{n\mu} j_{n\nu})})^\dagger
 \ket{0}_{\mu\nu}.
\eea
This last expression is written in terms of only $\wt{\cal U}$, and shows that
$\xi_F^{\mu\nu}$ is smooth.

\subsection{Action of each term in the Hamiltonian}

Basically terms $H_{(\mu)}$, $H_1$, $H_2$, $H_3$ and $H_4$ in $\hat{H}$
act on $\Psi$, $\chi_{L,E}$, $\Xi_B$ and $\Xi_F$ in $\hat{\psi}_{t,L}$ respectively, 
but derivative operators in them can act on other factors in $\hat{\psi}_{t,L}$.
So we shall investigate the action of each term carefully.
First let us consider $H_{(\mu)}\hat{\psi}_{t,L}$:
\bea
H_{(\mu)}\hat{\psi}_{t,L}
& = & 
\chi_{L,E}(\Lambda^M-tV^M, \Lambda^{IM},\theta^M_\alpha)\Bigg[
\Big[H_{(\mu)}\Psi\Big]\Xi_B\Xi_F
\nn & &
-\frac{1}{2}\Psi
\Big[\left(\frac{\p}{\p\lambda^{m_\mu}}\right)^2\Xi_B\Big]
\Xi_F
-\Big[\frac{\p}{\p\lambda^{m_\mu}}\Psi\Big]
\Big[\frac{\p}{\p\lambda^{m_\mu}}\Xi_B\Big]
\Xi_F
\nn & &
-\frac{1}{2}\Psi
\Xi_B
\Big[\left(\frac{\p}{\p\lambda^{m_\mu}}\right)^2\Xi_F\Big]
-\Big[\frac{\p}{\p\lambda^{m_\mu}}\Psi\Big]
\Xi_B
\Big[\frac{\p}{\p\lambda^{m_\mu}}\Xi_F\Big]
\nn & &
-\frac{1}{2}\Psi
\Xi_B
\Big[\left(\frac{\p}{\p\lambda^{Im_\mu}}\right)^2\Xi_F\Big]
-\Big[\frac{\p}{\p\lambda^{Im_\mu}}\Psi\Big]
\Xi_B
\Big[\frac{\p}{\p\lambda^{Im_\mu}}\Xi_F\Big]
\nn & &
-\frac{1}{2}\Psi
\Xi_B
\Big[\frac{\p}{\p Y_{(\mu)}^{(i_\mu j_\mu)}}\frac{\p}{\p Y_{(\mu)}^{(i_\mu j_\mu)}}\Xi_F\Big]
-\Big[\frac{\p}{\p Y_{(\mu)}^{(i_\mu j_\mu)}}\Psi\Big]
\Xi_B
\Big[\frac{\p}{\p Y_{(\mu)}^{(i_\mu j_\mu)}}\Xi_F\Big]
\nn & &
-\frac{1}{2}\Psi
\Xi_B
\Big[\frac{\p}{\p Y_{(\mu)}^{I(i_\mu j_\mu)}}\frac{\p}{\p Y_{(\mu)}^{I(i_\mu j_\mu)}}\Xi_F\Big]
-\Big[\frac{\p}{\p Y_{(\mu)}^{I(i_\mu j_\mu)}}\Psi\Big]
\Xi_B
\Big[\frac{\p}{\p Y_{(\mu)}^{I(i_\mu j_\mu)}}\Xi_F\Big]
\nn & &
-\Psi
\Big[\frac{\p}{\p\lambda^{m_\mu}}\Xi_B\Big]
\Big[\frac{\p}{\p\lambda^{m_\mu}}\Xi_F\Big]
\Bigg].
\eea
Since degrees of 
$\frac{\p}{\p\lambda^{m_\mu}}\Xi_B$,
$\left(\frac{\p}{\p\lambda^{m_\mu}}\right)^2\Xi_B$,
$\frac{\p}{\p\lambda^{m_\mu}}\Xi_F$,
$\left(\frac{\p}{\p\lambda^{m_\mu}}\right)^2\Xi_F$,
$\frac{\p}{\p\lambda^{Im_\mu}}\Xi_F$,
$\left(\frac{\p}{\p\lambda^{Im_\mu}}\right)^2\Xi_F$,
$\frac{\p}{\p Y_{(\mu)}^{(i_\mu j_\mu)}}\Xi_F$,
$\frac{\p}{\p Y_{(\mu)}^{(i_\mu j_\mu)}}\frac{\p}{\p Y_{(\mu)}^{(j_\mu i_\mu)}}\Xi_F$,
$\frac{\p}{\p Y_{(\mu)}^{I(i_\mu j_\mu)}}\Xi_F$, and
$\frac{\p}{\p Y_{(\mu)}^{I(i_\mu j_\mu)}}\frac{\p}{\p Y_{(\mu)}^{I(j_\mu i_\mu)}}\Xi_F$,
are $-1$, $-2$, $-1$, $-2$, $-2$, $-2$, $-1$, $-2$, $-2$, and $-2$ respectively,
\bea
H_{(\mu)}\hat{\psi}_{t,L}
& = & \chi_{L,E}(\Lambda^M-tV^M, \Lambda^{IM},\theta^M_\alpha)\Big[H_{(\mu)}\Psi\Big]\Xi_B\Xi_F
\nn & &
+\text{(terms of negative degree)}.
\eea
As is explained in Appendix B,
$H_{(\mu)}\Psi$ equals $E_{(\mu)}\Psi$ plus terms giving no contribution in
inner products in the limit $L\rightarrow\infty$.
Therefore, if we take large $L$ and $t$, the difference between
$H_{(\mu)}\hat{\psi}_{t,L}$ and $E_{(\mu)}\hat{\psi}_{t,L}$
can be made arbitrarily small in inner products.

Next let us consider $H_1\hat{\psi}_{t,L}$:
\bea
H_1\hat{\psi}_{t,L}
& = & 
\Big[H_1\chi_{L,E}\Big]
\Psi\Xi_B\Xi_F
\nn & &
-Q_{MN}\Bigg[
\frac{1}{2}\chi_{L,E}\Psi
\Big[\frac{\p}{\p\Lambda^M}\frac{\p}{\p\Lambda^N}\Xi_B\Big]
\Xi_F
+\Big[\frac{\p}{\p\Lambda^M}\chi_{L,E}\Big]\Psi
\Big[\frac{\p}{\p\Lambda^N}\Xi_B\Big]
\Xi_F
\nn & &
+\frac{1}{2}\chi_{L,E}\Psi\Xi_B
\Big[\frac{\p}{\p\Lambda^M}\frac{\p}{\p\Lambda^N}\Xi_F\Big]
+\Big[\frac{\p}{\p\Lambda^M}\chi_{L,E}\Big]
\Psi\Xi_B
\Big[\frac{\p}{\p\Lambda^N}\Xi_F\Big]
\nn & &
+\chi_{L,E}\Psi
\Big[\frac{\p}{\p\Lambda^M}\Xi_B\Big]
\Big[\frac{\p}{\p\Lambda^N}\Xi_F\Big]
\nn & &
+\frac{1}{2}\chi_{L,E}\Psi\Xi_B
\Big[\frac{\p}{\p\Lambda^{IM}}\frac{\p}{\p\Lambda^{IN}}\Xi_F\Big]
+\Big[\frac{\p}{\p\Lambda^{IM}}\chi_{L,E}\Big]
\Psi\Xi_B
\Big[\frac{\p}{\p\Lambda^{IN}}\Xi_F\Big]\Bigg].
\eea
Since degrees of 
$\frac{\p}{\p\Lambda^M}\Xi_B$,
$\frac{\p}{\p\Lambda^M}\frac{\p}{\p\Lambda^N}\Xi_B$,
$\frac{\p}{\p\Lambda^M}\Xi_F$,
$\frac{\p}{\p\Lambda^M}\frac{\p}{\p\Lambda^N}\Xi_F$,
$\frac{\p}{\p\Lambda^{IM}}\Xi_F$,
$\frac{\p}{\p\Lambda^{IM}}\frac{\p}{\p\Lambda^{IN}}\Xi_F$,
$\frac{\p}{\p\Lambda^M}\chi_{L,E}$, and
$\frac{\p}{\p\Lambda^{IM}}\chi_{L,E}$
are $-1$, $-2$, $-1$, $-2$, $-2$, $-2$, $0$, and $0$ respectively,
\beq
H_1\hat{\psi}_{t,L}=[H_1\chi_{L,E}]\Psi\Xi_B\Xi_F+\text{(terms of negative degree)}.
\eeq
As is explained in Appendix B,
$H_1\chi_{L,E}$ equals $E\chi_{L,E}$ plus terms giving no contribution in
inner products in the limit $L\rightarrow\infty$.
Therefore, if we take large $L$ and $t$, the difference between
$H_1\hat{\psi}_{t,L}$ and $E\hat{\psi}_{t,L}$
can be made arbitrarily small in inner products.

Action of $\sum_{\mu<\nu}(H_2^{\mu\nu}+H_3^{\mu\nu})$ on $\hat{\psi}_{t,L}$ is simple:
\beq
\sum_{\mu<\nu}(H_2^{\mu\nu}+H_3^{\mu\nu})\hat{\psi}_{t,L}
=\frac{1}{t}\Big[\sum_{\mu<\nu}(z^I_{\mu\nu})^2\Delta E^{\mu\nu}_{0F}(z_{\mu\nu}/t, (z^I_{\mu\nu})^2/t^2)
\Big]\hat{\psi}_{t,L}.
\eeq
Due to the factor $\chi_{(\mu)}$, nonzero contribution to inner products arises 
only when the arguments of $\Delta E^{\mu\nu}_{0F}$ are in some small
region of size $\sim\frac{L}{t}$. Let $R$ be a fixed region including this small region. Then
$|\Delta E^{\mu\nu}_{0F}|\leq \max_R|\Delta E^{\mu\nu}_{0F}|$,
and therefore the inner products containing $\sum_{\mu<\nu}(H_2^{\mu\nu}+H_3^{\mu\nu})\hat{\psi}_{t,L}$
are convergent and decay as $1/t$ or faster when $\frac{L}{t}\rightarrow 0$.

Let $H_4'$ be $H_4$ with the term containing $\hat{G}$ removed. Then
\bea
H_4\hat{\psi}_{t,L}
& = & H_4'\hat{\psi}_{t,L}
\nn & &
-\frac{1}{2}\mathop{{\sum}'}_{\mu,\nu}(z_{\mu\nu}^{-2})_{(i_\mu j_\nu)}{}^{(k_\mu l_\nu)}
\Big[
\hat{G}^F_{(k_\mu l_\nu)}\hat{G}^F_{(j_\nu i_\mu)}\hat{\psi}_{t,L}
+2\hat{G}^F_{(k_\mu l_\nu)}\hat{G}^B_{(j_\nu i_\mu)}\hat{\psi}_{t,L}
\nn & &
+\hat{G}^B_{(k_\mu l_\nu)}\hat{G}^B_{(j_\nu i_\mu)}\hat{\psi}_{t,L}
\Big],
\eea
We can easily see that $H'_4$ gives terms of negative degree, and $\hat{G}^F$ has no degree.
From the following expression of $\hat{G}^B$:
\bea
\hat{G}^B_{(i_\mu j_\nu)} & = & 
-(\Lambda^I_{(\mu)}-\Lambda^I_{(\nu)}+\lambda^I_{i_\mu}-\lambda^I_{j_\nu})\frac{\p}{\p Y^{I(i_\mu j_\nu)}}
\nn & &
-\sum_{k_\mu\not\neq i_\mu}Y_{(\mu)}^{I(k_\mu i_\mu)}\frac{\p}{\p Y^{I(k_\mu j_\nu)}}
+\sum_{l_\nu\not\neq j_\nu}Y_{(\nu)}^{I(j_\nu l_\nu)}\frac{\p}{\p Y^{I(i_\mu l_\nu)}}
\nn & &
+\sum_{\lambda\not\neq\mu,\nu}\Big[
Y^{I(j_\nu p_\lambda)}\frac{\p}{\p Y^{I(i_\mu p_\lambda)}}
-Y^{I(p_\lambda i_\mu)}\frac{\p}{\p Y^{I(p_\lambda j_\nu)}}
\Big]
\nn & &
+Y^{I(j_\nu i_\mu)}\Big[h_{(\mu)m_\mu}^{i_\mu}\frac{\p}{\p\lambda^{Im_\mu}}
-h_{(\nu)m_\nu}^{j_\nu}\frac{\p}{\p\lambda^{Im_\nu}}+
(h_{(0)M}^{\mu}-h_{(0)M}^{\nu})\frac{\p}{\p\Lambda^{IM}}
\Big]
\nn & &
+\sum_{k_\mu\not\neq i_\mu}Y^{I(j_\nu k_\mu)}\frac{\p}{\p Y_{(\mu)}^{I(i_\mu k_\mu)}}
-\sum_{l_\nu\not\neq j_\nu}Y^{I(l_\nu i_\mu)}\frac{\p}{\p Y_{(\nu)}^{I(l_\nu j_\nu)}},
\eea
we see that the degrees of $\hat{G}^B_{(j_\nu i_\mu)}\chi_{L,E}$, 
$\hat{G}^B_{(k_\mu l_\nu)}\hat{G}^B_{(j_\nu i_\mu)}\chi_{L,E}$,
$\hat{G}^B_{(j_\nu i_\mu)}\Psi$, 
$\hat{G}^B_{(k_\mu l_\nu)}\hat{G}^B_{(j_\nu i_\mu)}\Psi$,
$\hat{G}^B_{(j_\nu i_\mu)}\Xi_B$,
$\hat{G}^B_{(k_\mu l_\nu)}\hat{G}^B_{(j_\nu i_\mu)}\Xi_B$,
$\hat{G}^B_{(j_\nu i_\mu)}\Xi_F$, and
$\hat{G}^B_{(k_\mu l_\nu)}\hat{G}^B_{(j_\nu i_\mu)}\Xi_F$ are
$-1/2$, $-1$, $-1/2$, $-1$, $1/2$, $1$, $-5/2$, and $-5/2$ respectively.
The factor $z_{\mu\nu}^{-2}$ have degree $-2$.
Therefore $H_4\hat{\psi}_{t,L}$ consists of terms of negative degree.

In summary, $(\hat{H}-E-\sum_\mu E_{(\mu)})\hat{\psi}_{t,L}$ can be taken arbitrarily small 
in inner products if we take sufficiently large $L$ and $t$. 
This especially means the fact we want to show:
$||(\hat{H}-E-\sum_\mu E_{(\mu)})\hat{\psi}_{t,L}||\rightarrow 0~(L,t\rightarrow\infty)$,
and completes our proof of \eqref{theorem1}.

\section{Orthogonality of Trial Wavefunctions}

Let us take two wavefunctions $\psi_{t,L}$ and $\psi'_{t',L'}$ of the type we have
constructed in the previous section:
\bea
\hat{\psi}_{t,L} & = & \chi_{L,E}(\Lambda^M-tD^M, \Lambda^{IM},\theta^M_\alpha)
\prod_\mu\chi_{(\mu)L}\psi_{(\mu)}\prod_{\mu<\nu}\xi_B^{\mu\nu}\prod_{\mu<\nu}\xi_F^{\mu\nu},
\\
\hat{\psi}'_{t',L'} & = &
\chi'_{L',E'}(\Lambda^{\prime M'}-t'D^{M'}, \Lambda^{\prime IM'},\theta^{\prime M'}_\alpha)
\prod_{\mu'}\chi'_{(\mu')L'}\psi'_{(\mu')}\prod_{\mu'<\nu'}
\xi_B^{\prime\mu'\nu'}\prod_{\mu'<\nu'}\xi_F^{\prime\mu'\nu'},
\\
N & = & \sum_{\mu=1}^{n_b}N_\mu=\sum_{\mu'=1}^{n_b'}N'_{\mu'}.
\eea
Here and in the following, quantities related to $\hat{\psi}'_{t',L'}$ are denoted by primed symbols.
In this section we shall show that the inner product of $\psi_{t,L}$ and $\psi'_{t',L'}$
can be taken arbitrarily small:
for any positive real number $\epsilon$, there exists $L_0$ and $t_0$ such that
\beq
\forall L, L'>L_0~~\text{and}~~\forall t, t'>t_0,\quad
\big|\vev{\psi_{t,L},\psi'_{t',L'}}\big|<\epsilon,
\label{theorem2}
\eeq
where $L_0$ depends on $E$, $E'$ and $\epsilon$, and $t_0$ depends on $E$, $E'$, $\epsilon$, $L$ and $L'$.
This means that $\psi_{t,L}$ and $\psi'_{t',L'}$ give different branches
of the continuous spectrum.

First let us consider the case where both wavefunctions are based on the same partition of
$N\times N$ matrices i.e. $n_b=n_b'$, $N_\mu=N'_\mu$, but different eigenstates of $H_{(\mu)}$.
In this case, most part of both wavefunctions can be taken identical:
\beq
\chi_{(\mu)L}=\chi'_{(\mu)L},\quad
\xi_B^{\mu\nu}=\xi_B^{\prime\mu\nu},\quad \xi_F^{\mu\nu}=\xi_F^{\prime\mu\nu},
\eeq
and at least for one $\mu$, $\psi_{(\mu)}\not\neq\psi'_{(\mu)}$, and
$\vev{\psi_{(\mu)},\psi'_{(\mu)}}=0$.

In the limit $L\rightarrow\infty$, the difference between $\chi_{(\mu)L}\psi_{(\mu)}$
and $\psi_{(\mu)}$ is small, and therefore we expect that
\beq 
\vev{\chi_{(\mu)L}\psi_{(\mu)}, \chi'_{(\mu)L'}\psi'_{(\mu)}}\rightarrow
\vev{\psi_{(\mu)}, \psi'_{(\mu)}}\quad (L,L'\rightarrow\infty).
\eeq
A rigorous proof of this fact is given as follows (for notation see Appendix B.):
\bea
\vev{\chi_{(\mu)L}\psi_{(\mu)}, \chi'_{(\mu)L'}\psi'_{(\mu)}}
& = & 
\int\prod dx_{a_\mu}
\chi_{(\mu)L}^\dagger\psi_{(\mu)}^\dagger\chi'_{(\mu)L'}\psi'_{(\mu)}
\nn & = &
(A_{(\mu)L}A_{(\mu)L'})^{-1}\int_{S_\mu}\psi_{(\mu)}^\dagger\psi'_{(\mu)}
+\int_{R_\mu}\chi_{(\mu)L}^\dagger\psi_{(\mu)}^\dagger\chi'_{(\mu)L'}\psi'_{(\mu)},
\eea
where $R_\mu=\{x_{a_\mu}|\min(L,L')\leq r_\mu\}$ and $S_\mu=\{x_{a_\mu}|r_\mu\leq\min(L,L')\}$,
and we omit $\prod dx_{a_\mu}$ here 
and in the following.
The first term in the last expression of the above goes to $\vev{\psi_{(\mu)}, \psi'_{(\mu)}}$ as 
$L, L'\rightarrow\infty$. The second term can be rewritten as
\bea
& & \int_{R_\mu}
\chi_{(\mu)L}^\dagger\psi_{(\mu)}^\dagger\chi'_{(\mu)L'}\psi'_{(\mu)}
\nn & & 
=\frac{1}{2}\int_{R_\mu}\chi_{(\mu)L}^\dagger\chi'_{(\mu)L'}|\psi_{(\mu)}+\psi'_{(\mu)}|^2
-\frac{i}{2}\int_{R_\mu}\chi_{(\mu)L}^\dagger\chi'_{(\mu)L'}|\psi_{(\mu)}+i\psi'_{(\mu)}|^2
\nn & &
-\frac{1-i}{2}\int_{R_\mu}\chi_{(\mu)L}^\dagger\chi'_{(\mu)L'}|\psi_{(\mu)}|^2
-\frac{1-i}{2}\int_{R_\mu}\chi_{(\mu)L}^\dagger\chi'_{(\mu)L'}|\psi'_{(\mu)}|^2.
\eea
Each term of the above expression can be shown to go to zero as $L, L'\rightarrow \infty$.
For the first term,
\bea
0 & \leq & \int_{R_\mu}\chi_{(\mu)L}^\dagger\chi'_{(\mu)L'}|\psi_{(\mu)}+\psi'_{(\mu)}|^2
\nn & \leq & 
(A_{(\mu)L}A_{(\mu)L'})^{-1}\int_{R_\mu}|\psi_{(\mu)}+\psi'_{(\mu)}|^2
\nn & = & 
(A_{(\mu)L}A_{(\mu)L'})^{-1}\Big[
\int_{R_\mu}|\psi_{(\mu)}|^2+\int_{R_\mu}|\psi'_{(\mu)}|^2
+\int_{R_\mu}\psi_{(\mu)}^\dagger\psi'_{(\mu)}
+\int_{R_\mu}\psi_{(\mu)}^{\prime\dagger}\psi_{(\mu)}
\Big]
\nn & = & 
(A_{(\mu)L}A_{(\mu)L'})^{-1}\Big[
2+\vev{\psi_{(\mu)},\psi'_{(\mu)}}+\vev{\psi'_{(\mu)},\psi_{(\mu)}}
\nn & &
-\int_{S_\mu}|\psi_{(\mu)}|^2
-\int_{S_\mu}|\psi'_{(\mu)}|^2
-\int_{S_\mu}\psi_{(\mu)}^\dagger\psi'_{(\mu)}
-\int_{S_\mu}\psi_{(\mu)}^{\prime\dagger}\psi_{(\mu)}
\Big]
\nn & \rightarrow & 
0 \quad(L, L'\rightarrow \infty),
\eea
and similarly for the second term,
\bea
0 & \leq & \int_{R_\mu}\chi_{(\mu)L}^\dagger\chi'_{(\mu)L'}|\psi_{(\mu)}+i\psi'_{(\mu)}|^2
\nn & \leq & 
(A_{(\mu)L}A_{(\mu)L'})^{-1}\Big[
2+i\vev{\psi_{(\mu)},\psi'_{(\mu)}}-i\vev{\psi'_{(\mu)},\psi_{(\mu)}}
\nn & &
-\int_{S_\mu}|\psi_{(\mu)}|^2
-\int_{S_\mu}|\psi'_{(\mu)}|^2
-i\int_{S_\mu}\psi_{(\mu)}^\dagger\psi'_{(\mu)}
+i\int_{S_\mu}\psi_{(\mu)}^{\prime\dagger}\psi_{(\mu)}
\Big]
\nn & \rightarrow & 
0 \quad(L, L'\rightarrow \infty).
\eea
For the third term,
\bea
0 & \leq & \int_{R_\mu}\chi_{(\mu)L}^\dagger\chi'_{(\mu)L'}|\psi_{(\mu)}|^2
\nn & \leq & 
(A_{(\mu)L}A_{(\mu)L'})^{-1}\int_{R_\mu}|\psi_{(\mu)}|^2
\nn & = & 
(A_{(\mu)L}A_{(\mu)L'})^{-1}\Big[1-\int_{S_\mu}|\psi_{(\mu)}|^2
\Big]
\nn & \rightarrow & 
0 \quad(L, L'\rightarrow \infty).
\eea
Similarly, $\int_{R_\mu}\chi_{(\mu)L}^\dagger\chi'_{(\mu)L'}|\psi'_{(\mu)}|^2
\rightarrow 0 ~(L, L'\rightarrow \infty)$. 
Therefore $\int_{R_\mu}
\chi_{(\mu)L}^\dagger\psi_{(\mu)}^\dagger\chi'_{(\mu)L'}\psi'_{(\mu)}\rightarrow 0$,
and $\vev{\chi_{(\mu)L}\psi_{(\mu)}, \chi'_{(\mu)L'}\psi'_{(\mu)}}\rightarrow
\vev{\psi_{(\mu)}, \psi'_{(\mu)}}$.
Then
\bea
\vev{\hat{\psi}_{t,L},\hat{\psi}'_{t',L'}}
 & \rightarrow &
\vev{\chi_{L,E}(\Lambda^M-tD^M, \Lambda^{IM},\theta^M_\alpha),
\chi_{L',E'}(\Lambda^M-t'D^M, \Lambda^{IM},\theta^M_\alpha)}
\nn & &
\times\prod_\mu\vev{\psi_{(\mu)},\psi'_{(\mu)}}.
\eea
From the Cauchy-Schwarz inequality,
\beq
\Big|\vev{\chi_{L,E}(\Lambda^M-tD^M, \Lambda^{IM},\theta^M_\alpha),
\chi_{L',E'}(\Lambda^M-t'D^M, \Lambda^{IM},\theta^M_\alpha)}\Big|
\leq ||\chi_{L,E}||\cdot||\chi_{L',E'}||=1,
\eeq
and therefore $\vev{\hat{\psi}_{t,L},\hat{\psi}'_{t',L'}}\rightarrow 0$.

Next let us consider the case where the wavefunctions are based on different partitions of
$N\times N$ matrices. We assume that the first block of $\hat{\psi}'_{t',L'}$ is
larger than that of $\hat{\psi}_{t,L}$: $N'_1>N_1$
(If $N'_1=N_1$, we can just consider the second or later block.
If $N'_1<N_1$, we can just exchange $\hat{\psi}_{t,L}$ and $\hat{\psi}'_{t',L'}$.). 
The $\mu'=1$ block consists of the $\mu=1,2,\dots$ and $n$ blocks. The $\mu=n$ block
is not necessarily contained entirely by the $\mu'=1$ block.
Indices contained by both the $\mu'=1$ block and the $\mu=p$ block are denoted by $i_{p1}$.
Roughly speaking, $\hat{\psi}'_{t',L'}$ is nonzero only when eigenvalues of $X^9$ 
in the $\mu'=1$ block are near to each other, and $\hat{\psi}_{t,L}$ is
nonzero only when eigenvalues in the $\mu=1$ block and the $\mu=2$ block are
far from each other. So the supports of $\hat{\psi}_{t,L}$ and $\hat{\psi}'_{t',L'}$
have no intersection, and the inner product vanishes.

A rigorous proof of this fact is given as follows.
By the action of $K'$, $X'_{(1)}$ is block diagonalized into 
$\text{diag}(X_{(1)},X_{(2)},\dots,X_{(n-1)},\wt{X}_{(n)})$.
$\wt{X}_{(n)}$ may be equal to $X_{(n)}$ or part of $X_{(n)}$.
The inner product $\vev{\psi_{t,L},\psi'_{t',L'}}$ can be written in terms of 
the integral over $X_{(1)},X_{(2)},\dots,X_{(n-1)}$ and $\wt{X}_{(n)}$.
Then
\beq
Z_{(p)}^{(i_{p1}i_{p1})}=\Lambda_{(p)}+\lambda_{i_{p1}}=Z_{(1)}^{\prime\,(i_{p1}i_{p1})}
=\Lambda'_{(1)}+\lambda'_{i_{p1}},
\eeq
where $Z_{(1)}^{\prime\,(i_{p1}i_{p1})}$ are diagonal elements of 
$\text{diag}(X_{(1)},X_{(2)},\dots,X_{(n-1)},\wt{X}_{(n)})$, and
\beq
\sum_{m'_1}(\lambda^{\prime\,m'_1})^2
=\sum_{m'_1}h_{(1)m'_1}^{\prime~i'_1}h_{(1)m'_1}^{\prime~j'_1}\lambda'_{i'_1}\lambda'_{j'_1}
=\sum_{i'_1}(\lambda'_{i'_1})^2
\eeq
is evaluated as
\bea
\sum_{m'_1}(\lambda^{\prime\,m'_1})^2
 & = & \sum_{p=1}^{n}\sum_{i_{p1}}(\lambda'_{i_{p1}})^2
\nn & = &
\sum_{p=1}^{n}\sum_{i_{p1}}(\Lambda_{(p)}-\Lambda'_{(1)}+\lambda_{i_{p1}})^2
\nn & = &
\sum_{p=1}^{n}\sum_{i_{p1}}\big[(\Lambda_{(p)}-\Lambda'_{(1)})^2
+2(\Lambda_{(p)}-\Lambda'_{(1)})\lambda_{i_{p1}}
+(\lambda_{i_{p1}})^2\big]
\nn & \geq & 
\sum_{p=1}^{n}(\Lambda_{(p)}-\Lambda'_{(1)})^2
+2(\Lambda_{(n)}-\Lambda'_{(1)})\sum_{i_{n1}}\lambda_{i_{n1}}.
\label{lp2ineq}
\eea
In the support of $\psi_{t,L}$, we can restrict the range of
$\lambda^{m_n}$ to $|\lambda^{m_n}|\leq L+\ell<2L$. Since in general $\sum a_i\geq-\sum|a_i|$,
\bea
(\Lambda_{(n)}-\Lambda'_{(1)})\sum_{i_{n1}}\lambda_{i_{n1}}
& \geq &
-\big|\Lambda_{(n)}-\Lambda'_{(1)}\big|\sum_{i_{n1}}|h_{(n)m_n}^{i_{n1}}||\lambda^{m_n}|
\nn & > &
-2L\big|\Lambda_{(n)}-\Lambda'_{(1)}\big|\sum_{i_{n1}}|h_{(n)m_n}^{i_{n1}}|.
\eea
Using 
\bea
\Lambda_{(p)}-\Lambda'_{(1)} & = &
\left(\frac{\sum_\mu \mu N_\mu}{N}-p\right)t
-\left(\frac{\sum_{\mu'} \mu' N'_{\mu'}}{N}-1\right)t'
\nn & &
+\frac{h_{(0)M}^p}{N_p}W_{MN}\wt{\Lambda}^N
-\frac{h_{(0)M'}^{\prime\,1}}{N'_1}W'_{M'N'}\wt{\Lambda}^{\prime\,N'},
\eea
and noting that $|\wt{\Lambda}^M|\leq L$ and $|\wt{\Lambda}^{\prime\,M'}|\leq L'$
in the support of $\psi_{t,L}$ and $\psi'_{t',L'}$, 
the second term of the last line in \eqref{lp2ineq} is evaluated as
\bea
(\Lambda_{(n)}-\Lambda'_{(1)})\sum_{i_{n1}}\lambda_{i_{n1}}
 & > &
-2L\sum_{i_{n1}}|h_{(n)m_n}^{i_{n1}}|
\Bigg[
\left|\frac{\sum_\mu \mu N_\mu}{N}-n\right|t
+\left(\frac{\sum_{\mu'} \mu' N'_{\mu'}}{N}-1\right)t'
\nn & &
+L\sum_N\left|\frac{h_{(0)M}^n}{N_n}W_{MN}\right|
 +L'\sum_{N'}\left|\frac{h_{(0)M'}^{\prime\,1}}{N'_1}W'_{M'N'}\right|
\Bigg].
\eea
Similarly, the first term of the last line in \eqref{lp2ineq} is evaluated as
\bea
\sum_{p=1}^{n}(\Lambda_{(p)}-\Lambda'_{(1)})^2
 & = & \sum_{p=1}^{n}\Bigg[
\left(\frac{\sum_\mu \mu N_\mu}{N}-p\right)t
-\left(\frac{\sum_{\mu'} \mu' N'_{\mu'}}{N}-1\right)t'
\nn & &
+\frac{h_{(0)M}^p}{N_p}W_{MN}\wt{\Lambda}^N
-\frac{h_{(0)M'}^{\prime\,1}}{N'_1}W'_{M'N'}\wt{\Lambda}^{\prime\,N'}
\Bigg]^2
\nn & = &
n\left[
\left(\frac{\sum_\mu \mu N_\mu}{N}-\frac{n+1}{2}\right)t
-\left(\frac{\sum_{\mu'} \mu' N'_{\mu'}}{N}-1\right)t'\right]^2
+\frac{n}{12}(n^2-1)t^2
\nn & &
+2\Bigg[\sum_{p=1}^{n}
\left(\frac{\sum_\mu \mu N_\mu}{N}-p\right)\frac{h_{(0)M}^p}{N_p}W_{MN}\wt{\Lambda}^N
\nn & &
-n\left(\frac{\sum_\mu \mu N_\mu}{N}-\frac{n+1}{2}\right)
 \frac{h_{(0)M'}^{\prime\,1}}{N'_1}W'_{M'N'}\wt{\Lambda}^{\prime\,N'}
\Bigg]t
\nn & &
-2\left(\frac{\sum_{\mu'} \mu' N'_{\mu'}}{N}-1\right)
\Bigg[\sum_{p=1}^{n}\frac{h_{(0)M}^p}{N_p}W_{MN}\wt{\Lambda}^N
-n\frac{h_{(0)M'}^{\prime\,1}}{N'_1}W'_{M'N'}\wt{\Lambda}^{\prime\,N'}
\Bigg]t'
\nn & &
+\sum_{p=1}^{n}\Bigg[
\frac{h_{(0)M}^p}{N_p}W_{MN}\wt{\Lambda}^N
-\frac{h_{(0)M'}^{\prime\,1}}{N'_1}W'_{M'N'}\wt{\Lambda}^{\prime\,M'}
\Bigg]^2
\nn & \geq &
n\left[
\left(\frac{\sum_\mu \mu N_\mu}{N}-\frac{n+1}{2}\right)t
-\left(\frac{\sum_{\mu'} \mu' N'_{\mu'}}{N}-1\right)t'\right]^2
+\frac{n}{12}(n^2-1)t^2
\nn & &
-2\Bigg[L\sum_N\left|
\sum_{p=1}^{n}\left(\frac{\sum_\mu \mu N_\mu}{N}-p\right)\frac{h_{(0)M}^p}{N_p}W_{MN}\right|
\nn & & 
+nL'\left|\frac{\sum_\mu \mu N_\mu}{N}-\frac{n+1}{2}\right|
 \sum_{N'}\left|\frac{h_{(0)M'}^{\prime\,1}}{N'_1}W'_{M'N'}\right|
\Bigg]t
\nn & &
-2\left(\frac{\sum_{\mu'} \mu' N'_{\mu'}}{N}-1\right)
\nn & & \times
\Bigg[L\sum_N\left|\sum_{p=1}^{n}\frac{h_{(0)M}^p}{N_p}W_{MN}\right|
+nL'\sum_{N'}\left|\frac{h_{(0)M'}^{\prime\,1}}{N'_1}W'_{M'N'}\right|
\Bigg]t'.
\eea
Therefore
\bea
\sum_{m_1}(\lambda^{\prime\,m_1})^2 & > &
n\left[
\left(\frac{\sum_\mu \mu N_\mu}{N}-\frac{n+1}{2}\right)t
-\left(\frac{\sum_{\mu'} \mu' N'_{\mu'}}{N}-1\right)t'
+J_1\right]^2
\nn & &
+\frac{n}{12}(n^2-1)\left(t-\frac{12}{n(n^2-1)}J_2\right)^2-J_3,
\label{lm2ineq}
\eea
where
\bea
J_1 & = & \frac{L}{n}\sum_N\left|\sum_{p=1}^n\frac{h_{(0)M}^p}{N_p}W_{MN}\right|
 +L'\sum_{N'}\left|\frac{h_{(0)M'}^{\prime\,1}}{N'_1}W'_{M'N'}\right|
 +\frac{L}{n}\sum_{i_{n1}}|h_{(n)m_n}^{i_{n1}}|,
\\
J_2 & = & nL'\left(\frac{\sum_\mu \mu N_\mu}{N}-\frac{n+1}{2}
+\left|\frac{\sum_\mu \mu N_\mu}{N}-\frac{n+1}{2}\right|
\right)\sum_{N'}\left|\frac{h_{(0)M'}^{\prime\,1}}{N'_1}W'_{M'N'}\right|
\nn & & +
L\left(\frac{\sum_\mu \mu N_\mu}{N}-\frac{n+1}{2}\right)
\sum_N\left|\sum_{p=1}^n\frac{h_{(0)M}^p}{N_p}W_{MN}\right|
\nn & & +
L\sum_N\left|\sum_{p=1}^n\frac{h_{(0)M}^p}{N_p}W_{MN}
\left(\frac{\sum_\mu \mu N_\mu}{N}-p\right)
\right|
\nn & &
+L\left(
\left|\frac{\sum_\mu \mu N_\mu}{N}-n\right|
 +\frac{\sum_\mu \mu N_\mu}{N}-\frac{n+1}{2}\right)
 \sum_{i_{n1}}|h_{(n)m_n}^{i_{n1}}|,
\\
J_3 & = & 2L\left(L\sum_N\left|\frac{h_{(0)M}^n}{N_n}W_{MN}\right|
 +L'\sum_{N'}\left|\frac{h_{(0)M'}^{\prime\,1}}{N'_1}W'_{M'N'}\right|\right)
 \sum_{i_{n1}}|h_{(n)m_n}^{i_{n1}}|
\nn & &
+nJ_1^2+\frac{12}{n(n^2-1)}J_2^2.
\eea
Note that $J_1$ and $J_2$ are sums of terms proportional to $L$ or $L'$, and 
$J_3$ is a sum of terms proportional to $L^2$, $LL'$ or $L^{\prime 2}$.
If we take $t$ and $t'$ sufficiently larger than $L$ and $L'$, 
the right hand side of \eqref{lm2ineq} becomes larger than $(L'+\ell)^2$.
On the other hand, in the support of $\psi'_{t',L'}$, 
$\sum_{m_1}(\lambda^{\prime\,m_1})^2\leq (r'_1)^2\leq(L'+\ell)^2$. 
Thus we see that for $t$ and $t'$ sufficiently larger than $L$ and $L'$,
the supports of $\psi_{t,L}$ and $\psi'_{t',L'}$ have no intersection,
and $\vev{\psi_{t,L},\psi'_{t',L'}}=0$. This completes our proof of \eqref{theorem2}.

\section{Discussion}

We have constructed a trial wavefunction $\psi_{t,L}$ for each partition of $N$
and each choice of normalizable bound states $\psi_{(\mu)}$, which shows continuous 
energy spectrum. We have shown that the wavefunctions for different partitions of $N$ 
or choices of $\psi_{(\mu)}$ are orthogonal to each other in the limit
$t, L\rightarrow\infty$.

Note that we regarded the same partitions of $N$ in different orders 
as different. For example, if $N=N_1+N_2=N'_1+N'_2$ and $N_1=N'_2\not\neq N_2=N'_1$,
$\{N_1,N_2\}$ and $\{N'_1,N'_2\}$ are regarded as different partitions.
However wavefunctions corresponding to these partitions stand for essentially 
the same branch of the continuous spectrum, and different only in the positions of
the bound states.

In our trial wavefunctions, the centers of the bound states $\psi_{(\mu)}$
in $X^I$ directions are at the origin. We can shift the positions of those centers
to $\Lambda_0^{IM}$ by shifting the arguments of $\chi_{L,E}$ as
$\chi_{L,E}(\Lambda^M-tD^M, \Lambda^{IM}-\Lambda_0^{IM},\theta^M_\alpha)$.
Though computations are a little more complicated, we can show that the same 
propositions as \eqref{theorem1} and \eqref{theorem2} also hold in this case.

The bounds $L_0$ and $t_0$ depend on the partitions of $N$
and the choices of normalizable bound states $\psi_{(\mu)}$.
Since the number of the partitions of $N$ is finite,
$L_0$ and $t_0$ can be taken independent of the partitions of $N$
just by taking the maximum of $L_0$ and $t_0$ for various partitions.
Similarly, if the number of normalizable states is finite,
$L_0$ and $t_0$ can be taken independent of the choices of $\psi_{(\mu)}$.
However if there exist infinitely many normalizable bound states,
it is not clear if $L_0$ and $t_0$ can be taken independent of them.

Though our construction is in the supermembrane matrix model obtained from $(9+1)\,$D SYM, 
similar construction can be done in matrix models obtained from lower dimensional SYM.

Our analysis supports the intuition about the structure of the energy spectrum of the 
SU($N$) supermembrane matrix quantum mechanics.
However we still have no proof that there is no branch other than the types
we have constructed. 

\vs{.5cm}
\noindent
{\large\bf Acknowledgments}\\[.2cm]
I would like to thank Y.~Isokawa, H.~Kita and Y.~Kuramoto for helpful discussions.

\renewcommand{\theequation}{\Alph{section}.\arabic{equation}}
\appendix
\addcontentsline{toc}{section}{Appendix}
\vs{.5cm}
\noindent
{\Large\bf Appendix}
\section{SU($N$) Lie algebra}
\label{appa}
\setcounter{equation}{0}

A Cartan subalgebra $\{h_m;\;m=1,2,\dots,N-1\}$
of SU($N$) Lie algebra can be given as a set of diagonal traceless $N\times N$ matrices.
For example,
\beq
h_m=\text{diag}(h_m^1, h_m^2,\dots, h_m^N)
=\frac{1}{\sqrt{m(m+1)}}\left(\begin{array}{@{\,}ccccc@{\,}}
\begin{array}{ccc}
1 & & \\
& \ddots & \\
& & 1 \\
\end{array} & \Bigg\}~m & & & \\
 & -m & & & \\
 & & 0 & & \\
 & & & \ddots & \\
 & & & & 0
\end{array}\right).
\eeq
These satisfy $\text{tr}(h_mh_n)=\delta_{mn}$. Then
a Cartan-Weyl basis is given by
\beq
\{h_m,E_{ij};\;m=1,2,\dots,N-1,\; i,j=1,2,\dots,N,\;i\not\neq j\},
\eeq
where $E_{ij}$ is the matrix whose only nonzero component is at $i$-th row and $j$-th column: $(E_{ij})^i{}_j=1$.
Commutation relations of these operators are
\beq
{}[h_m,h_n]=0,\quad [h_m,E_{ij}]=(h^i_m-h^j_m)E_{ij},\quad
[E_{ij},E_{kl}]=\delta_{jk}E_{il}-\delta_{il}E_{kj},
\label{suncom}
\eeq
Note that $\sum_{i=1}^Nh^i_m=0$,
and the right hand side of the last equation in \eqref{suncom} can contain $h_m$, because $E_{ii}$ is 
diagonal and can be rewritten in terms of $h_m$.

An ordinary hermitian basis $\{T^a;\;a=1,2,\dots,N^2-1\}$ satisfying $(T^a)^\dagger=T^a$ and 
$\text{tr}(T^aT^b)=\delta^{ab}$
is given by
\beq
\{T^a;\;a=1,2,\dots,N^2-1\}=\{h_m,E_{ij}^+,E_{ij}^-;\;
m=1,2,\dots,N-1,\; 1\leq i<j\leq N\},
\eeq
where
\beq
E_{ij}^+=\frac{1}{\sqrt{2}}(E_{ij}+E_{ji}),\quad E_{ij}^-=\frac{i}{\sqrt{2}}(E_{ij}-E_{ji}).
\eeq
Due to the following:
\beq
\sum_{i,j}(E_{ij})^k{}_l(E_{ji})^{k'}{}_{l'}=\delta^k{}_{l'}\delta^{k'}{}_l,\quad
\sum_{m=1}^{N-1}(h_m)^i{}_j(h_m)^{i'}{}_{j'}=\delta^i{}_j\delta^{i'}{}_{j'}\left(\delta^{ii'}-\frac{1}{N}\right).
\eeq
$T^a$ satisfy 
$\sum_a(T^a)^i{}_j(T^a)^{i'}{}_{j'}=\delta^i{}_{j'}\delta^{i'}{}_j-\frac{1}{N}\delta^i{}_j\delta^{i'}{}_{j'}$.
An element $\theta^aT^a$ of SU($N$) Lie algebra can be expanded in various ways. We define
$\theta^m$, $\theta^{(\pm ij)}$ and $\theta^{(ij)}$ as follows:
\beq
\theta^aT^a = \sum_{m=1}^{N-1}\theta^mh_m+\sum_{i<j}(\theta^{(+ij)}E_{ij}^++\theta^{(-ij)}E_{ij}^-)
 = \sum_{m=1}^{N-1}\theta^mh_m+\sum_{i\not\neq j}\theta^{(ij)}E_{ij}.
\eeq

\section{Properties of auxiliary functions}
\label{appb}
\setcounter{equation}{0}

First we define the following functions of class $C^\infty$:
\beq
f_0(x)= \begin{cases} e^{-1/x} & (x>0) \\ 0 & (x\leq 0) \end{cases},\quad
f_1(x)=f_0(x)f_0(1-x),\quad
f_2(x)=\int_{-\infty}^xdyf_1(y),
\eeq
\beq
F(x)=f_1(x)/[f_2(1)]^{1/2},\quad G(x)=f_2(x)/f_2(1).
\eeq
Then the following functions
\beq
F_L(x)=\frac{1}{\sqrt{2L}}F\left(\frac{x+L}{2L}\right),\quad
G_L(x)=G\left(\frac{L+\ell-x}{\ell}\right)
\eeq
have the profiles shown in Figure 1 and 2.
$\ell$ can be arbitrary positive real number and is fixed throughout the analysis in this paper.
The support of $F_L(x)$ is $-L\leq x\leq L$, and $F_L(x)$ is normalized: 
$\int_{-\infty}^\infty dx(F_L(x))^2=1$.
$G_L(x)$ is considered only in the region $x\geq 0$, and $G_L(x)=1$ for $0\leq x\leq L$.
Its support is $0\leq x\leq L+\ell$.
\begin{figure}[H]
\begin{minipage}{80mm}
\begin{figure}[H]
\centering
\includegraphics*[width=60mm,clip]{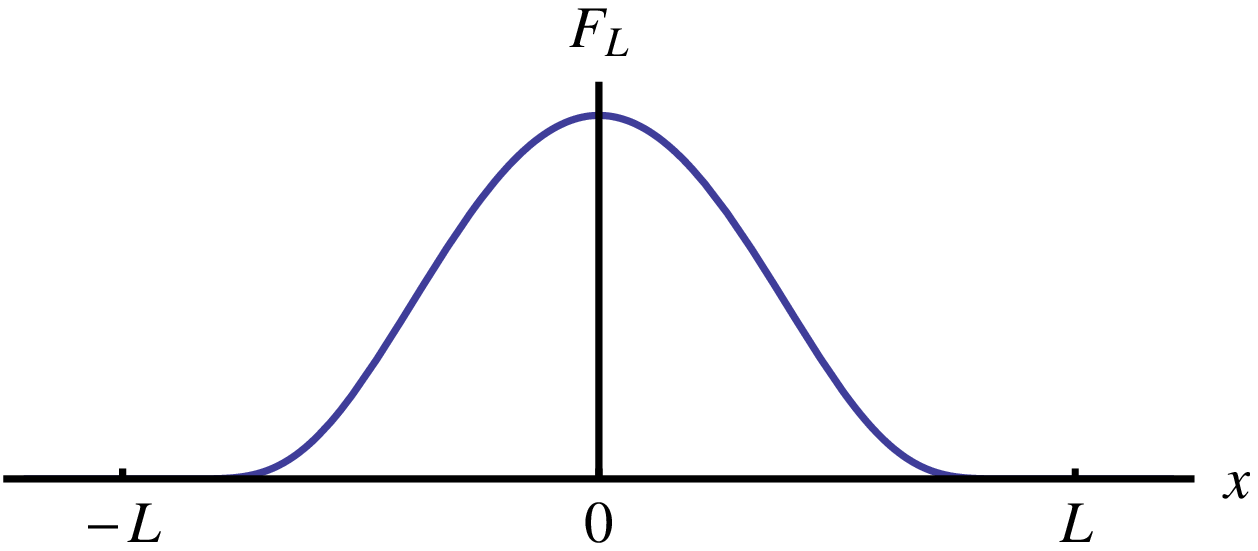}
\caption{$F_L(x)$}
\end{figure}
\end{minipage}
\begin{minipage}{80mm}
\begin{figure}[H]
\centering
\includegraphics*[width=60mm,clip]{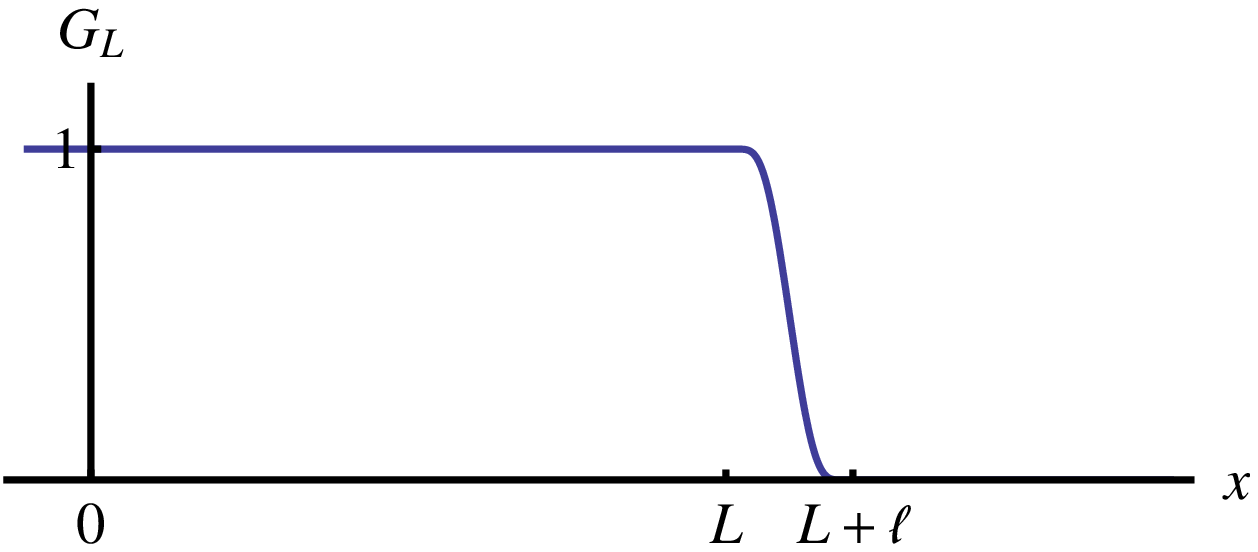}
\caption{$G_L(x)$}
\end{figure}
\end{minipage}
\end{figure}
The absolute maxima of $F_L(x)$ and its derivatives $F_L^{(n)}(x)$ are proportional to $L^{-(n+1/2)}$, and
the absolute maxima of $G_L(x)$ and its derivatives $G_L^{(n)}(x)$ are independent of $L$:
\bea
\max |F_L^{(n)}(x)| & = & \frac{1}{(2L)^{n+1/2}}\max |F^{(n)}(x)|
 \equiv \frac{1}{L^{n+1/2}}f_{(n)},
\\
\max |G_L^{(n)}(x)| & = & \frac{1}{\ell^n}\max |G^{(n)}(x)|
 \equiv g_{(n)}.
\eea
Variables $\lambda^{m_\mu}$, $\lambda^{Im_\mu}$, $Y_{(\mu)}^{(i_\mu j_\mu)}$ 
and $Y_{(\mu)}^{I(i_\mu j_\mu)}$ are collectively
denoted by $x_{a_\mu}$, and let $r_\mu$ be
\bea
r_\mu & = & \sqrt{\sum_{a_\mu}(x_{a_\mu})^2}
\nn & = & \sqrt{\sum(\lambda^{m_\mu})^2+\sum(\lambda^{Im_\mu})^2
 +\sum Y_{(\mu)}^{(i_\mu j_\mu)}Y_{(\mu)}^{(j_\mu i_\mu)}
 +\sum Y_{(\mu)}^{I(i_\mu j_\mu)}Y_{(\mu)}^{I(j_\mu i_\mu)}}
\nn & = & 
\sqrt{\text{tr}(X_{(\mu)}-\Lambda_{(\mu)}I_{(\mu)})^2
+\text{tr}(X^I_{(\mu)}-\Lambda^I_{(\mu)}I_{(\mu)})^2}.
\eea
$r_\mu$ is invariant under $K$. Take a normalizable energy eigenstate $\psi_{(\mu)}(x_{a_\mu})$
of SU($N_{(\mu)}$) subsystem with the eigenvalue $E_{(\mu)}$. It is normalized:
\beq
||\psi_{(\mu)}||^2=\int\prod dx_{a_\mu}\psi_{(\mu)}^\dagger\psi_{(\mu)}=1,
\eeq
and it is assumed that inner products of derivatives of $\psi_{(\mu)}$ are finite. For instance,
\beq
\vev{\psi_{(\mu)},\p_{x_{a_\mu}}\psi_{(\mu)}}
 =\int\prod dx_{a_\mu}\psi_{(\mu)}^\dagger\p_{x_{a_\mu}}\psi_{(\mu)}
\eeq
is finite. The integral
\beq
A_{(\mu)L}^2=\int\prod dx_{a_\mu}(G_L(r_\mu))^2\psi_{(\mu)}^\dagger\psi_{(\mu)}
\eeq
is positive, is less than 1, monotonically increases as $L$ increases, and goes to 1
as $L\rightarrow\infty$. $\chi_{(\mu)L}(r_\mu)$ is defined so that
$||\chi_{(\mu)L}\psi_{(\mu)}||=1$:
\beq
\chi_{(\mu)L}(r_\mu)\equiv G_L(r_\mu)/A_{(\mu)L}.
\eeq
For $N_\mu=1$ we define $\chi_{(\mu)L}$ as $\chi_{(\mu)L}=1$.
To show that the diffenrence between $H_{(\mu)}(\chi_{(\mu)L}\psi_{(\mu)})$
and $E_{(\mu)}(\chi_{(\mu)L}\psi_{(\mu)})$ can be made arbitrarily small by taking large $L$,
we first show that inner products of $\chi_{(\mu)L}\psi_{(\mu)}$ with some $\chi_{(\mu)L}$
replaced by their derivatives go to zero as $L\rightarrow \infty$. For example,
\bea
\big|\vev{\chi_{(\mu)L}\psi_{(\mu)}, (\p_{x_{a_\mu}}\chi_{(\mu)L})\psi_{(\mu)}}\big|
 & = & A_{(\mu)L}^{-2}\Big|\int_{L\leq r_\mu\leq L+\ell}\prod dx_{b_\mu}G_L(r_\mu)
\frac{x_{a_\mu}}{r_\mu}\p_{r_\mu}G_{L,\ell}(r_\mu)\psi_{(\mu)}^\dagger\psi_{(\mu)}\Big|
\nn & < & 
A_{(\mu)L}^{-2}g_{(1)}\int_{L\leq r_\mu\leq L+\ell}\prod dx_{b_\mu}\psi_{(\mu)}^\dagger\psi_{(\mu)}
\nn & \leq & 
A_{(\mu)L}^{-2}g_{(1)}\int_{L\leq r_\mu}\prod dx_{b_\mu}\psi_{(\mu)}^\dagger\psi_{(\mu)}
\nn & = & 
A_{(\mu)L}^{-2}g_{(1)}\Big(1-\int_{r_\mu\leq L}\prod dx_{b_\mu}\psi_{(\mu)}^\dagger\psi_{(\mu)}\Big)
\nn & \rightarrow & 0 \quad(L\rightarrow \infty).
\eea
Similarly, inner products containing second derivatives of $\chi_{(\mu)L}$ 
can be shown to go to zero as $L\rightarrow \infty$. Then
\bea
\vev{\chi_{(\mu)L}\psi_{(\mu)},
[H_{(\mu)}-E_{(\mu)}](\chi_{(\mu)L}\psi_{(\mu)})}
& = & \vev{\chi_{(\mu)L}\psi_{(\mu)},
\chi_{(\mu)L}[H_{(\mu)}-E_{(\mu)}]\psi_{(\mu)}}
\nn & &
-\vev{\chi_{(\mu)L}\psi_{(\mu)},
\frac{\p}{\p x_{a_\mu}}\chi_{(\mu)L}
\frac{\p}{\p x_{a_\mu}}\psi_{(\mu)}}
\nn & &
-\frac{1}{2}\vev{\chi_{(\mu)L}\psi_{(\mu)},
\frac{\p^2}{\p x_{a_\mu}^2}\chi_{(\mu)L}\psi_{(\mu)}}
\nn & \rightarrow &
0 \quad(L\rightarrow \infty).
\eea
Similarly, $\vev{[H_{(\mu)}-E_{(\mu)}](\chi_{(\mu)L}\psi_{(\mu)}),
[H_{(\mu)}-E_{(\mu)}](\chi_{(\mu)L}\psi_{(\mu)})}\rightarrow 0
\quad(L\rightarrow \infty)$.

Next we define (the bosonic part of) $\chi_{L,E}$.
$Q_{MN}=h_{(0)M}^\mu{\cal N}_{\mu\nu}h_{(0)N}^\nu$ is real and symmetric, and therefore
is diagonalizable by an orthogonal matrix $W_{MN}$: $Q_{MN}=q_LW_{ML}W_{NL}$.
The eigenvalues $q_M$ are real.
Noting that $Q_{MN}$ satisfies 
$(Q^n)_{MN}=h_{(0)M}^\mu({\cal N}^n)_{\mu\nu}h_{(0)N}^\nu$, 
its eigenvalue equation $0=\det(Q-qI)\equiv w(q)$ is evaluated as
\beq
w(q)=-\frac{1}{N}\left(\sum_\mu\frac{N_\mu}{N_\mu-q}\right)\prod_\nu(N_\nu-q).
\eeq
Since $w(q)<0$ for $q\leq 0$, all the eigenvalues $q_M$ are positive.

Let $\Lambda^a$ denote $\Lambda^NW_{NM}$ and $\Lambda^{IN}W_{NM}$.
Then $H_1=-\frac{1}{2}q_a\frac{\p}{\p\Lambda^a}\frac{\p}{\p\Lambda^a}$, 
and $\chi_{L,E}$ is defined as
\beq
\chi_{L,E}(\Lambda^a)\equiv\prod_bF_{L}(\Lambda^b)e^{ik_b\Lambda^b},
\eeq
where $k_a$ are arbitrary real numbers satisfying $\sum_aq_ak_ak_a=2E$.
$q_a$ is defined as $q_M$ for $a=M$ and $a=IM$.
$\chi_{L,E}$ is normalized:
\beq
||\chi_{L,E}||^2=\int_{-\infty}^\infty \prod_ad\Lambda^a|\chi_{L,E}(\Lambda^a)|^2=1.
\eeq
Note that $\prod_Md\Lambda^M\prod_{I,N}d\Lambda^{IN}=\prod_ad\Lambda^a$.
We can show that the difference between $H_1\chi_{L,E}$
and $E\chi_{L,E}$ can be made arbitrarily small by taking large $L$. For example,
\bea
\big|\vev{\chi_{L,E}, [H_1-E]\chi_{L,E}}\big| & = &
\frac{1}{2}\left|\int_{-\infty}^\infty\prod_ad\Lambda^a\chi_{L,E}^\dagger
q_b\left(-\frac{d}{d\Lambda^b}\frac{d}{d\Lambda^b}-k_bk_b\right)\chi_{L,E}\right|
\nn & = & 
\frac{1}{2}\left|\int_{-L}^L dx F_L(x)\left(-\sum_aq_a\frac{d^2}{dx^2}-2iq_bk_b\frac{d}{dx}\right)F_L(x)\right|
\nn
& < & \frac{1}{2}\int_{-L}^L dx \frac{f_{(0)}}{L^{1/2}}
\left(\Big|\sum_aq_a\Big|\frac{f_{(2)}}{L^{5/2}}+2|q_ak_a|\frac{f_{(1)}}{L^{3/2}}\right)
\nn
& < & f_{(0)}\left(\Big|\sum_aq_a\Big|\frac{f_{(2)}}{L^2}+2|q_ak_a|\frac{f_{(1)}}{L}\right)
\nn
& \rightarrow & 0 \quad(L\rightarrow \infty).
\eea
Similarly, $\vev{[H_1-E]\chi_{L,E}, [H_1-E]\chi_{L,E}}\rightarrow 0
\quad(L\rightarrow \infty)$.

\section{Smoothness of eigenvalues and unitary transformations}
\label{appc}
\setcounter{equation}{0}

Eigenvalues of matrices are determined by solving the eigenvalue equations.
Let us consider an $N\times N$ matrix $A=(a_{ij})$, and let its eigenvalues be $y_i$.
Then the eigenvalue equation 
\beq
0=F(y,\bm{x})\equiv y^N+x_1y^{N-1}+x_2y^{N-2}+\dots+x_N=(y-y_1(\bm{x}))(y-y_2(\bm{x}))\dots (y-y_N(\bm{x}))
\label{appc:eeq}
\eeq 
is an algebraic equation of degree $N$, and its coefficients $x_i$, or $\bm{x}$ collectively, are
smooth functions of $a_{ij}$. Since, in general, solutions to algebraic equations are continuous functions of
coefficients of the equations, $y_i=y_i(\bm{x})$ are continuous functions of $\bm{x}$, or $a_{ij}$.
However $y_i=y_i(\bm{x})$ are not always differentiable, as can be easily seen from 
explicit expressions of solutions to equations of lower degrees.
By applying the standard implicit function theorem to \eqref{appc:eeq},
we see that if $y_i(\bm{x}_0)$ is not equal to $y_j(\bm{x}_0)$ for any $j\not\neq i$, 
$y_i(\bm{x})$ is smooth in the neighborhood of $\bm{x}_0$. When the equation has multiple solutions
$y_i(\bm{x}_0)=y_j(\bm{x}_0)$, the implicit function theorem cannot be applied at $\bm{x}=\bm{x}_0$
because of $\p_yF(y_i(\bm{x}_0),\bm{x}_0)=0$. 

Even when we have multiple solutions, we can show that the sum of those solutions are smooth: 
If $y_1(\bm{x}_0)=y_2(\bm{x}_0)=\dots =y_n(\bm{x}_0)$ and $y_1(\bm{x}_0)\not\neq y_i(\bm{x}_0)$ for $i\geq n$,
$y_1(\bm{x})+y_2(\bm{x})+\dots +y_n(\bm{x})$ is smooth at $\bm{x}_0$.
In addition,  $[y_1(\bm{x})]^q+[y_2(\bm{x})]^q+\dots +[y_n(\bm{x})]^q$ are also smooth
for any positive integer $q$. This can be shown as follows: 
We regard the function $F(y,\bm{x})$ as one on the complex plane $y\in\mathbb{C}$, then the following holds
in the neighborhood of $\bm{x}_0$.
\beq
[y_1(\bm{x})]^q+[y_2(\bm{x})]^q+\dots +[y_n(\bm{x})]^q
=\oint_{C(\bm{x}_0)}\frac{dy}{2\pi i}\frac{y^q\p_yF(y,\bm{x})}{F(y,\bm{x})},
\label{appc:sumsol}
\eeq
where the fixed contour $C(\bm{x}_0)$ encircles $y=y_1(\bm{x}_0), y_2(\bm{x}_0), \dots$ and $y_n(\bm{x}_0)$.
From the continuity of $y_i(\bm{x})$, $y=y_1(\bm{x}), y_2(\bm{x}), \dots$ and $y_n(\bm{x})$
are on the inside of $C(\bm{x}_0)$, and $y=y_{n+1}(\bm{x}), y_{n+2}(\bm{x}), \dots$ and $y_N(\bm{x})$ are
on the outside of $C(\bm{x}_0)$ for any $\bm{x}$ near $\bm{x}_0$.
Clearly the right hand side of \eqref{appc:sumsol} is differentiable at $\bm{x}_0$ with respect to $\bm{x}$.

If $A=(a_{ij})$ is hermitian, it is diagonalizable by a unitary matrix $u$:
$uAu^{-1}=\text{diag}(y_1,y_2,\dots,y_N)$, and eigenvalues $y_i$ are real.
Although $y_i$ are continuous functions of $a_{ij}$, elements of $u$ are not even continuous
at the points where some of the eigenvalues degenerate.
Let us see this in detail. $u$ is constructed from normalized eigenvectors $\bm{v}_i$ as follows:
\beq
u^\dagger=(\bm{v}_1,\bm{v}_2,\dots,\bm{v}_N),\quad
A\bm{v}_i=y_i\bm{v}_i,\quad \bm{v}^\dagger_i\cdot\bm{v}_i=1,
\eeq
and when $y_i$ is not a degenerate eigenvalue,
\beq
\bm{v}_i\propto\begin{pmatrix} \wt{\Delta}^{(i)}_{j1} \\ \wt{\Delta}^{(i)}_{j2} \\
 \vdots \\ \wt{\Delta}^{(i)}_{jN} \end{pmatrix},\quad \wt{\Delta}^{(i)}_{kl}=(k,l)~\text{cofactor of}~A-y_iI,
\eeq
for any $j$. Since the rank of $A-y_iI$ is $N-1$,
at least one of $\wt{\Delta}^{(i)}_{kl}$ is nonzero. So we can choose $j$ such that $\bm{v}_i$ is a
nonzero vector. Then $\bm{v}_i$ is given by
\beq
\bm{v}_i=\frac{1}{\sqrt{\sum_k|\wt{\Delta}^{(i)}_{jk}|^2}}
\begin{pmatrix} \wt{\Delta}^{(i)}_{j1} \\ \wt{\Delta}^{(i)}_{j2} \\
 \vdots \\ \wt{\Delta}^{(i)}_{jN} \end{pmatrix}.
\eeq
Since $\wt{\Delta}^{(i)}_{kl}$ are smooth functions of $a_{ij}$, so is $\bm{v}_i$.
However, when $y_i$ becomes degenerate, all of $\wt{\Delta}^{(i)}_{kl}$ vanish, and
the above expression of $\bm{v}_i$ is not well-defined. (We can take a limit into the point
where $y_i$ is degenerate, but the limit depends on how we approach the point.)

If we consider $A$ only in a region $R$, where $R$ is such that $A$ always has a
degenerate eigenvalue throughout the region, we can find smooth orthonormal eigenvectors.
($R$ usually has nonzero codimension in the entire space of $a_{ij}$.)
Let $A$ have $r$-fold eigenvalue $0$ in $R$, and let other eigenvalues never be zero in $R$.
Then the rank of $A$ is $N-r$, and there exists nonzero $(N-r)\times(N-r)$ cofactor.
if
\beq
\Delta\equiv\det\begin{pmatrix} a_{11} & a_{12} & \dots & a_{1,N-r} \\
a_{21} & a_{22} & \dots & a_{2,N-r} \\
\vdots & & \ddots & \vdots \\
a_{N-r,1} & a_{N-r,2} & \dots & a_{N-r,N-r} \end{pmatrix}
\eeq
is nonzero, then $r$ eigenvectors $\bm{v}_1,\dots,\bm{v}_r$ corresponding to the eigenvalue $0$ are 
\beq
\bm{v}_1\propto\begin{pmatrix} \Delta_{11} \\ \Delta_{21} \\ \vdots \\ \Delta_{N-r,1} \\
 -\Delta \\ 0 \\ 0 \\ \vdots \\ 0 \end{pmatrix},
\quad
\bm{v}_2\propto\begin{pmatrix} \Delta_{12} \\ \Delta_{22} \\ \vdots \\ \Delta_{N-r,2} \\
 0 \\ -\Delta \\ 0 \\ \vdots \\ 0 \end{pmatrix},
\quad\dots,\quad
\bm{v}_r\propto\begin{pmatrix} \Delta_{1r} \\ \Delta_{2r} \\ \vdots \\ \Delta_{N-r,r} \\
 0 \\ 0 \\ \vdots \\ 0 \\ -\Delta \end{pmatrix},
\eeq
where $\Delta_{ij}$ is $\Delta$ with $i$-th column replaced
by $(a_{1,N-r+j},a_{2,N-r+j},\dots,a_{N-r,N-r+j})^T$.
$\Delta_{ij}$ and $\Delta$ are smooth functions of $a_{ij}$, and
$\bm{v}_i$ can be smoothly orthonormalized by Gram-Schmidt process where $\Delta$ is nonzero.
If $\Delta=0$, we can replace $\Delta$ by other nonzero $(N-r)\times(N-r)$ cofactors and
can construct $r$ smooth eigenvectors similarly to the above.
When two $(N-r)\times(N-r)$ cofactors are nonzero, we can construct two sets of $r$
smooth orthonormal eigenvectors $\{\bm{e}_1,\dots,\bm{e}_r\}$ and $\{\bm{f}_1,\dots,\bm{f}_r\}$,
and these vectors are related to each other by the smooth unitary matrix
$V_{ji}=\bm{f}_j^\dagger\cdot\bm{e}_i$: $\bm{e}_i=\bm{f}_jV_{ji}$.
Therefore, if $A$ goes out of the region where $\Delta\not\neq 0$ into another region
where another $(N-r)\times(N-r)$ cofactor is nonzero as we change the elements 
of $A$ continuously, we can use the "transition matrix" $V_{ji}$  
to keep the smoothness of the eigenvectors. 
(This is similar to the situation we encounter 
when we go from one coordinate patch to another on a manifold.)

Using the above fact, we can show that if two sets of eigenvalues 
$\{y_1,y_2,\dots,y_r\}$ and $\{y_{r+1},y_{r+2},\dots,y_N\}$
have no intersection, then there exist a smooth unitary matrix $U$, a smooth hermitian $r\times r$
matrix $A_1$, and a smooth hermitian $(N-r)\times(N-r)$ matrix $A_2$ such that
\beq
UAU^{-1}=\begin{pmatrix} A_1 & 0 \\ 0 & A_2 \end{pmatrix},
\label{appc:blockdiag}
\eeq
and, $A_1$ and $A_2$ have eigenvalues $\{y_1,y_2,\dots,y_r\}$ and $\{y_{r+1},y_{r+2},\dots,y_N\}$
respectively.

This can be shown by using the following hermitian matrix:
\beq
{\cal A}\equiv (A-y_1I)(A-y_2I)\dots(A-y_rI)
=\sum_{n=0}^r (-1)^nP_n A^{r-n},
\eeq
where $P_0=1$ and $P_m=\sum_{1\leq i_1<i_2<\dots<i_m\leq r} y_{i_1}y_{i_2}\dots y_{i_m}$.
$P_m$ can be expressed in terms of $\sum_{i=1}^r(y_i)^q$, which we have already shown to be smooth.
Therefore elements of ${\cal A}$ are also smooth.
${\cal A}$ has $r$-fold degenerate eigenvalue $0$, and we have already shown that we can construct
$r$ smooth orthonormal eigenvectors $\bm{e}_1, \bm{e}_2,\dots,\bm{e}_r$ which span the union of
the eigenspaces of $A$ corresponding to eigenvalues $\{y_1,y_2,\dots,y_r\}$. Similarly, by considering 
${\cal A}'\equiv(A-y_{r+1}I)(A-y_{r+2}I)\dots(A-y_NI)$, we can construct smooth orthonormal vectors
$\bm{e}'_1,\bm{e}'_2,\dots,\bm{e}'_{N-r}$ which span the union of the eigenspaces of $A$
corresponding to eigenvalues $\{y_{r+1},y_{r+2},\dots,y_N\}$. 
These two sets of vectors are orthogonal to each other.
Therefore
\beq
U^\dagger=(\bm{e}_1, \bm{e}_2,\dots,\bm{e}_r,\bm{e}'_1,\bm{e}'_2,\dots,\bm{e}'_{N-r})
\eeq
is unitary and smooth, and satisfies \eqref{appc:blockdiag}.
When we have to go out of the region where $\bm{e}_i$ and  $\bm{e}'_i$ are well-defined,
we can use the "transition matrix" of the block diagonal form to keep the smoothness of
$U$, $A_1$ and $A_2$:
\beq
U^\dagger\rightarrow U^\dagger\begin{pmatrix} V_1^\dagger & 0 \\ 0 & V_2^\dagger \end{pmatrix}
\label{transition}
\eeq
i.e. if we parametrize the space of $a_{ij}$ by $A_1$, $A_2$ and $U$, we need some coordinate
patches, and those patches are connected by the transition matrices.

By applying this fact repeatedly, $A$ can be block diagonalized into smaller hermitian matrices
$A_i$ by a smooth unitary matrix, corresponding to disjoint sets of eigenvalues:
\beq
UAU^{-1}=A_D\equiv\begin{pmatrix}
A_1 & & & \\ & A_2 & & \\
 & & \ddots & \\ & & & A_n
\end{pmatrix}.
\label{appc:blockdiag2}
\eeq
Each $A_i$ can further be diagonalized by multiplying appropriate 
block diagonal unitary matrix $u$ to \eqref{appc:blockdiag2}:
\beq
uUAU^{-1}u^{-1}=\begin{pmatrix}
y_1 & & & \\ & y_2 & & \\
 & & \ddots & \\ & & & y_N
\end{pmatrix},\quad
u=\begin{pmatrix}
U_1 & & & \\ & U_2 & & \\
 & & \ddots & \\ & & & U_n
\end{pmatrix},
\eeq
but these $U_i$ are not necessarily smooth.

If we have a smooth function $\hat{f}(A_D)$ of $A_D$ which is invariant under the action of 
block diagonal unitary matrices $u$:
$\hat{f}(A_D)=\hat{f}(uA_Du^{-1})$, then it can be extended to a smooth function 
$f(A)=\hat{f}(A_D)$ of $A=U^{-1}A_DU$ which is invariant under the action of unitary matrices:
$f(A)=f(U'AU^{\prime -1})$. Note that due to the invariance $\hat{f}(A_D)=\hat{f}(uA_Du^{-1})$,
the transition matrix \eqref{transition} can act on $A_D$ consistently.

\newcommand{\J}[4]{{\sl #1} {\bf #2} (#3) #4}
\newcommand{\andJ}[3]{{\bf #1} (#2) #3}
\newcommand{\AP}{Ann.\ Phys.\ (N.Y.)}
\newcommand{\MPL}{Mod.\ Phys.\ Lett.}
\newcommand{\NP}{Nucl.\ Phys.}
\newcommand{\PL}{Phys.\ Lett.}
\newcommand{\PR}{Phys.\ Rev.}
\newcommand{\PRL}{Phys.\ Rev.\ Lett.}
\newcommand{\PTP}{Prog.\ Theor.\ Phys.}
\newcommand{\hepth}[1]{{\tt hep-th/#1}}
\newcommand{\arxivhep}[1]{{\tt arXiv.org:#1 [hep-th]}}

\end{document}